\documentclass{aastex63}
\usepackage[caption=false]{subfig}
\usepackage{graphicx}
\usepackage{multirow}
\usepackage{graphicx}
\usepackage{booktabs}
\usepackage{makecell}
\usepackage{bm}
\usepackage{amsmath,amssymb}
\usepackage[T1]{fontenc}
\graphicspath{{./}{figures/}}
\usepackage{longtable}
\usepackage{epstopdf}
\usepackage[]{amsmath}
\usepackage{threeparttablex}
\usepackage{cases}

\begin{document}

\title{Investigating the Correlation between ZTF Tidal Disruption Events and IceCube High-energy Neutrinos }
\correspondingauthor{Yun-Feng Liang, Xiang-Gao Wang}
\email{E-mail: liangyf@gxu.edu.cn, wangxg@gxu.edu.cn}

\author{Ming-Xuan Lu}
\affiliation{Guangxi Key Laboratory for Relativistic Astrophysics,
	School of Physical Science and Technology, Guangxi University, Nanning 530004,
	China}
\affiliation{GXU-NAOC Center for Astrophysics and Space Sciences, Nanning 530004, People's Republic of China}
\author{Yun-Feng Liang}
\affiliation{Guangxi Key Laboratory for Relativistic Astrophysics,
	School of Physical Science and Technology, Guangxi University, Nanning 530004,
	China}
\affiliation{GXU-NAOC Center for Astrophysics and Space Sciences, Nanning 530004, People's Republic of China}
\author{Xiang-Gao Wang}
\affiliation{Guangxi Key Laboratory for Relativistic Astrophysics,
	School of Physical Science and Technology, Guangxi University, Nanning 530004,
	China}
\affiliation{GXU-NAOC Center for Astrophysics and Space Sciences, Nanning 530004, People's Republic of China}
\author{Xue-Rui Ouyang}
\affiliation{Guangxi Key Laboratory for Relativistic Astrophysics,
	School of Physical Science and Technology, Guangxi University, Nanning 530004,
	China}
\affiliation{GXU-NAOC Center for Astrophysics and Space Sciences, Nanning 530004, People's Republic of China}


\begin{abstract}
Investigating the correlation between the TDE population and IceCube neutrinos could help us better understand whether TDEs could be potential high-energy neutrino emitters. In this paper, we perform a systematic search for TDEs that are associated with neutrinos in a sample including 143 IceCube neutrino alert events and 52 TDEs classified by the Zwicky Transient Facility (ZTF) - Bright Transient Survey (BTS). Furthermore, considering that the TDEs/TDE candidates reported as potential IceCube neutrino emitters are all accompanied by infrared (IR) echo emissions, we further select the TDEs with IR echoes from these 52 TDEs as a subsample to examine the correlation with neutrinos. Based on the Wide-field Infrared Survey Explorer (WISE) mission database, seven TDEs are identified as having IR echoes.
Then we employ Monte Carlo simulations to quantify the correlation between the TDE sample/subsample and IceCube neutrinos. Finally, after considering spatial and temporal criteria, the seven TDEs with IR echoes show the most significant correlation at a 2.46$\sigma$ confidence level. 
If we tentatively further take the time delay factor into account, the correlation enhances to a 2.66$\sigma$ confidence level.
The correlation is primarily contributed by two TDEs: AT2019dsg and AT2019azh. The latter's association with a neutrino alert, IC230217A, is newly reported in this work.
We discussed the possible physical connection between AT2019azh and the neutrino event IC230217A.
\end{abstract}

\keywords{Neutrino astronomy (1100) --- Tidal disruption (1696) --- High energy astrophysics (739)}

%
\section{Introduction}           
\label{sect:intro}
The detection of the TeV-PeV diffuse astrophysical neutrino flux by IceCube \citep{Aartsen:2013jdh, IceCube:2014stg, Aartsen:2015knd, IceCube:2016umi, Aartsen:2020aqd, Abbasi:2020jmh, IceCube:2021uhz} has unveiled new possibilities for exploration in both astrophysics 
\citep{Kistler:2006hp, Beacom:2007yu, Murase:2010cu, Murase:2013ffa, Murase:2013rfa, Ahlers:2014ioa, Tamborra:2014xia, Murase:2014foa, Bechtol:2015uqb, Kistler:2016ask, Bartos:2016wud, Sudoh:2018ana, Bartos:2018jco, Bustamante:2019sdb, Hovatta:2020lor, Ackermann:2022rqc} 
and particle physics \citep{GonzalezGarcia:2005xw, Ioka:2014kca, Ng:2014pca, Aartsen:2017kpd, Bustamante:2017xuy,Zhou:2019vxt, Zhou:2019frk, Zhou:2021xuh, IceCube:2021rpz, Ackermann:2022rqc, Guo:2023axz, Plestid:2024bva, Lu:2024jbq}. 
Despite a significant number of studies have been conducted to identify the sources of these astrophysical neutrinos \citep{Abbasi:2010rd,Aartsen:2013uuv,Aartsen:2014cva,ANTARES:2015moa,IceCube:2016tpw,IceCube:2017der,Aartsen:2018ywr,Aartsen:2019fau,IceCube:2019lzm,IceCube:2020svz, Zhou:2021rhl,Chang:2022hqj, Kouch:2024xtd}, it remains unclear which astrophysical sources are the primary origins of these neutrinos, and only a very small number of sources (e.g., the Seyfert galaxy NGC 1068) have been identified as neutrino sources \citep{IceCube:2022der}. 

Among various astrophysical sources, tidal disruption events (TDEs) have been proposed as potential emitters of high-energy neutrinos. TDE is a class of bright transients, and 
the general picture of TDE involves a star being tidally disrupted when it approaches sufficiently close to a supermassive black hole (SMBH). A portion of the disrupted stellar debris is subsequently accreted by the SMBH, resulting in the production of an electromagnetic flare that occurs over timescales ranging from months to years \citep{1988Natur.333..523R}.
Numerous models have been proposed to explain the origin of the nonthermal electromagnetic and neutrino emissions from TDEs, including relativistic jets \citep{Wang:2011ip,Wang:2015mmh,Dai:2016gtz,Senno:2016bso,Lunardini:2016xwi,Liu:2020isi,Hayasaki:2021jem}, accretion disks \citep{Hayasaki:2019kjy}, wide-angle outflows/hidden winds \citep{Fang:2020bkm}, and tidal stream interactions \citep{Dai:2015eua,Hayasaki:2019kjy}. 

Among the TDEs/TDE candidates that have been reported as potential IceCube neutrino emitters, three notable events include: one identified TDE AT2019dsg \citep{2021NatAs...5..510S}, and two TDE candidates, AT2019fdr \citep{Reusch:2021ztx} and AT2019aalc \citep{vanVelzen:2021zsm}, which correspond to the possibly associated alert events IC191001A, IC200530A, and IC191119A, respectively.
In these three events, AT2019dsg has been classified as a member of TDE by the Zwicky Transient Facility (ZTF). The long-lived nonthermal emission detected in AT2019dsg \citep{Cendes:2021bvp,2021NatAs...5..510S} suggests that mildly relativistic outflows may provide an ideal environment for neutrino production. In contrast, the classification of AT2019fdr as a TDE is somewhat uncertain, as the flare occurred in a known active galactic nucleus (AGN), complicating its origin due to the presence of an accretion disk \citep[e.g.,][]{Trakhtenbrot:2019seb,Frederick:2020sjz,Cendes:2021bvp}. 
Similarly, AT2019aalc is also posited to be a flare originating from another known AGN \citep{vanVelzen:2021zsm,Veres:2024qcm}. These three events share several comparable characteristics; for instance, they have bright optical luminosities, accompanying infrared (IR) echoes, i.e., delayed IR rising after the discovery date of TDE. Further, all of these events have a time delay for the neutrino detection, which has been explained by some models,
for instance, protons generated by the delayed choked jet precession \citep{Mukhopadhyay:2023mld,Lu:2023miv}, the slow acceleration of protons by the plasma/magnetic reconnection \citep{Murase:2020lnu}, 
an increase of the neutrino production efficiency over time assuming that the radius of neutrino production evolves similarly to that of the black body \citep{Winter:2020ptf},
and the protons accelerated by the outflow/jet interacting with protons/photons after traveling to the cloud/dust region \citep{Wu:2021vaw,Winter:2022fpf}.

Besides these three sources, there are other TDEs / TDE candidates that have also been reported as possible IceCube high-energy neutrino emitters. For instance, \citet{Jiang:2023kbb} reported two obscured TDE candidates as potential neutrino emitters based on a sample of mid-IR outbursts in nearby galaxies (MIRONG). \citet{Yuan:2024foi} reported a super-bright TDE candidate with IR emission that satisfies spatial-temporal coincidence when considering systematic errors. Recently, \citet{Li:2024qcp} reported another TDE that is associated with a potential high-energy neutrino flare at a 2.9$\sigma$ confidence level, which also accompanies IR echoes.

As more and more TDEs/TDE candidates are reported as potential IceCube high-energy neutrino emitters, investigating the correlation between the TDE population and high-energy neutrinos is helpful for understanding whether the TDE population could indeed be contributors to the IceCube neutrino flux. 
In this work, we investigate the hypothesis that the TDE population is correlated with IceCube alert events\footnote{\url{https://gcn.gsfc.nasa.gov/amon_icecube_gold_bronze_events.html}}. 
We utilize the TDEs classified by ZTF - Bright Transient Survey (BTS)\footnote{\url{https://sites.astro.caltech.edu/ztf/bts/bts.php}} \citep{Perley:2020ajb}, which is a public catalog with a strict pipeline to conduct a classification for transient sources, and a magnitude-limited ($m < 19\,{\rm mag}$ in either the $g$ or $r$ filter) survey for extragalactic transients in the ZTF public stream \citep{Perley:2020ajb}. 
We systematically search for spatial association events between TDEs of the ZTF-BTS catalog and IceCube alert events, while allowing a time delay between TDEs and the alert events. 
Since IR echoes are detected from all the TDEs/TDE candidates that are reported as potential IceCube neutrino emitters, we also systematically search for the IR emission at the coordinates of ZTF-BTS TDEs using the Wide-field Infrared Survey Explorer (WISE) mission database \citep{2010AJ....140.1868W}. 
Thus, we further adopt TDEs with IR echoes as a subsample to investigate the correlation with IceCube high-energy neutrinos.

\section{Sample}           
\subsection{TDE sample}
\label{sect:TDE sample}

\begin{figure*}
\center
\includegraphics[width=0.6\textwidth]{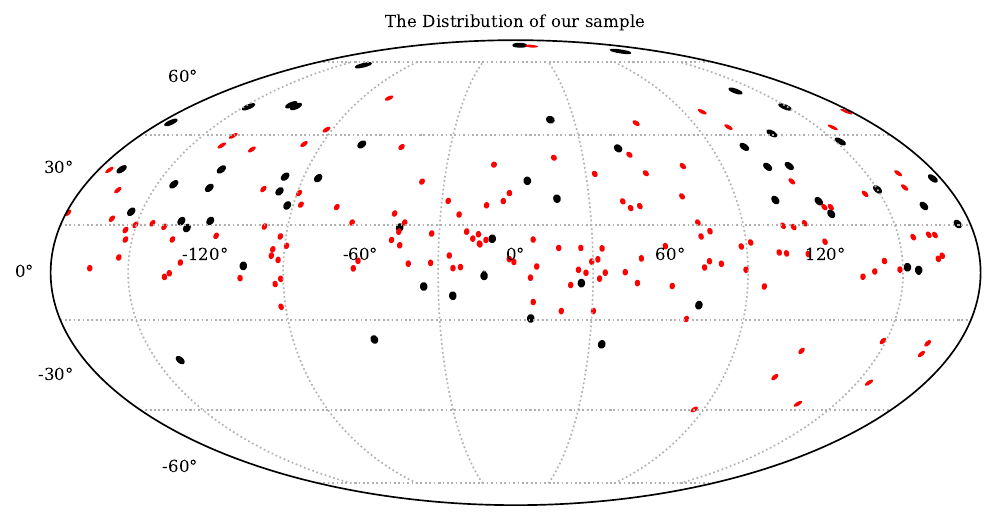}
\caption{Sky distributions in celestial coordinates of the 52 TDEs and 143 alert events considered in this work. The black points represent the positions of the TDEs. The red points represent the locations of alert events.}
\label{fig:TDE sample}
\end{figure*}

Since 2018, the ZTF has conducted a three-day cadence survey of the visible northern sky ($\sim 3\pi$), announcing newly found transient candidates via public alerts \citep{Masci:2018neq,Graham:2019qsw,2019PASP..131a8002B,Fremling:2019dvl}.
In December 2020, the ZTF public survey has transitioned to Phase-II operations, with the survey shifting to a two-day cadence.
To complement the photometric survey, the ZTF-BTS undertakes an extensive spectroscopic campaign, aiming to spectroscopically classify all extragalactic transients with a peak magnitude of $<18.5$ mag in the $g$ or $r$ filters (transients down to 19.0 mag are also included when spectroscopic resources are available), with the classifications made publicly available \citep{Fremling:2019dvl}. The survey contains over 10,000 objects and is updated nearly daily$^2$.
This catalog offers a comprehensive and purely magnitude-limited sample of extragalactic transients, suitable for detailed statistical and demographic analysis. 
The ZTF-BTS catalog sets a series of stringent quality criteria to classify transients, such as magnitude limit, observation times and Galactic extinction limit (see \citet{Perley:2020ajb} for details). 
Therefore, this source catalog can provide some TDEs/candidates with relatively reliable classification as our sample for performing the correlation analysis. We select TDEs according to the classifications reported by ZTF-BTS, including 61 TDEs up to November 2024\footnote{Note that AT2019fdr is classified as a supernova and AT2019aalc is not included in the ZTF-BTS catalog.}.

Then we search for the IR echoes from coordinates of ZTF-BTS TDEs using the full Near-Earth Object WISE Reactivation database\footnote{\url{https://irsa.ipac.caltech.edu/cgi-bin/Gator/nph-dd?catalog=neowiser_p1bs_psd}} \citep[NEOWISE-R;][]{2014ApJ...792...30M,https://doi.org/10.26131/irsa144}. The full NEOWISE-R database is up to July 2024. {Most parts of the sky are visited every six months, receiving approximately 10 observations during a 24 h period of each visit \citep{2010AJ....140.1868W}.}
To ensure that the IR echoes of TDEs can be included in the NEOWISE-R database, we limit our TDE sample to before January 2024 ({i.e., six months before the end of the NEOWISE-R data}), due to the fact that the IR echoes of TDEs often have significant time delay {and considering the six-month survey cadence of NEOWISE}. After the time cut, there are 52 TDEs remaining in our sample, as shown in Fig.~\ref{fig:TDE sample}. The NEOWISE-R database is based on an imaging survey centered at 3.4 and 4.6 $\mu$m (labeled W1 and W2). The photometric measurements are obtained from the fitting of the point-spread function (PSF), where the PSFs are estimated from observations involving tens of thousands of stars. The NEOWISE-R magnitudes are measured through PSF fitting for individual exposures, which have been utilized to examine the IR flares for astrophysical sources \citep[e.g.,][]{2021ApJS..252...32J}.

\begin{figure*}
\center
\includegraphics[width=0.3\textwidth]{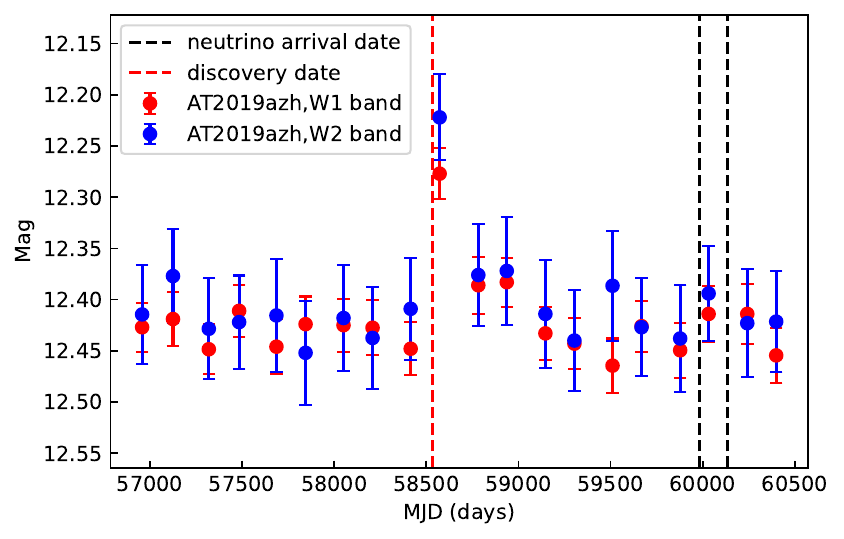}
\includegraphics[width=0.3\textwidth]{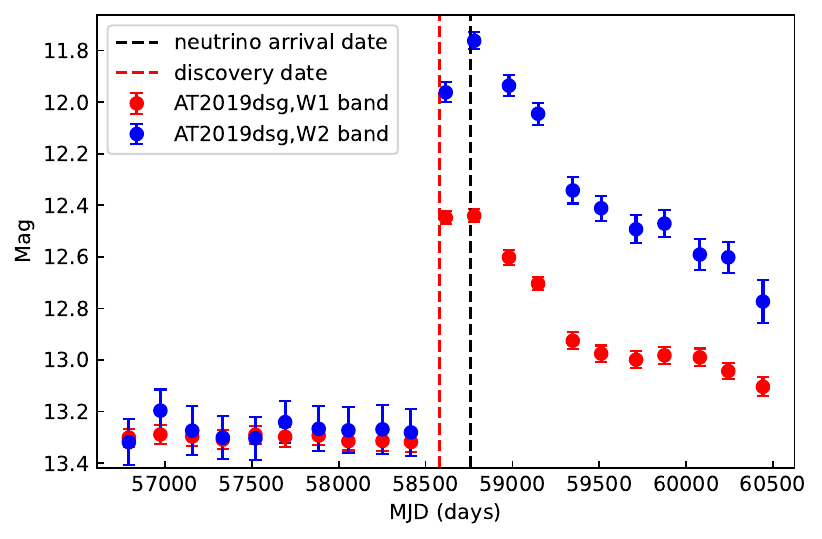}
\includegraphics[width=0.3\textwidth]{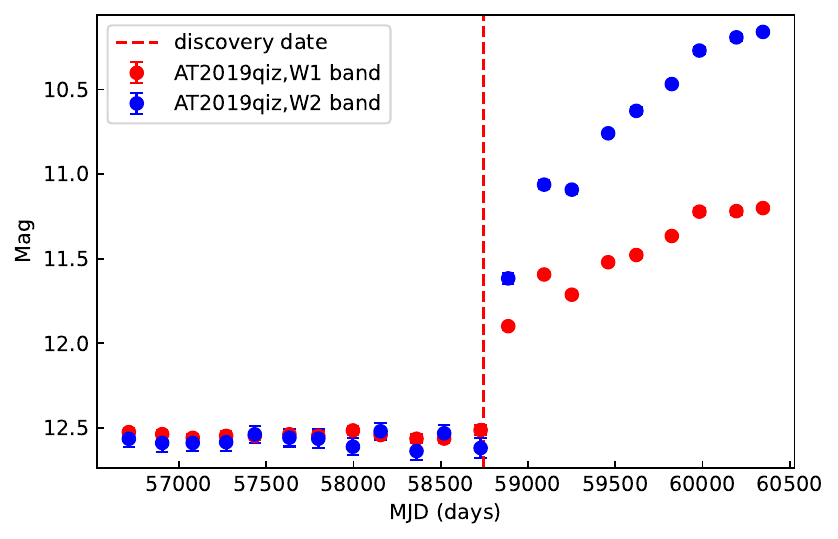}
\includegraphics[width=0.3\textwidth]{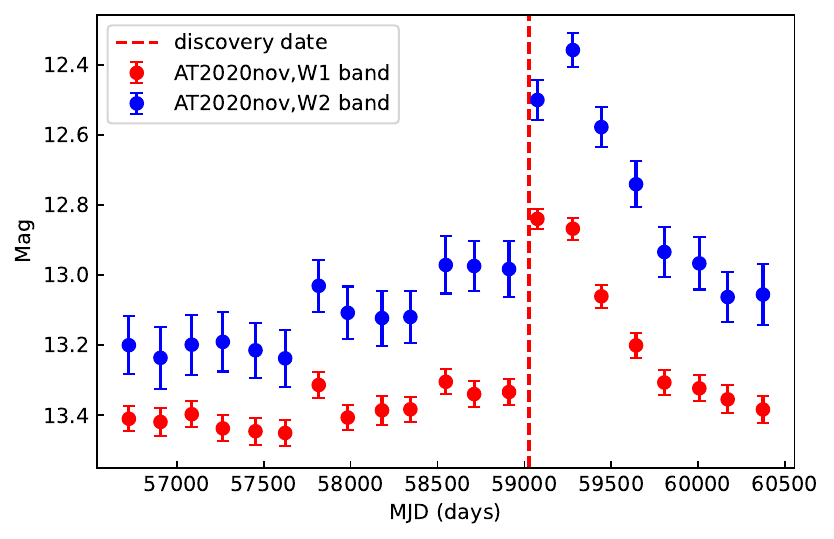}
\includegraphics[width=0.3\textwidth]{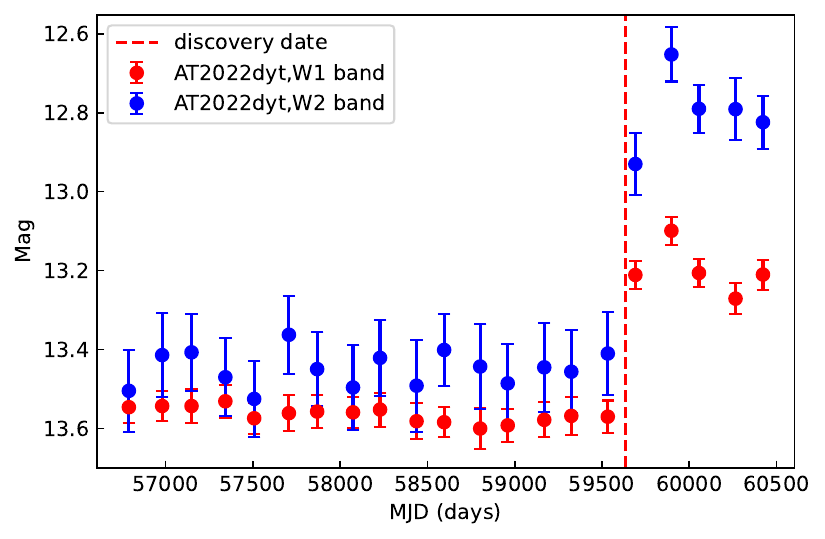}
\includegraphics[width=0.3\textwidth]{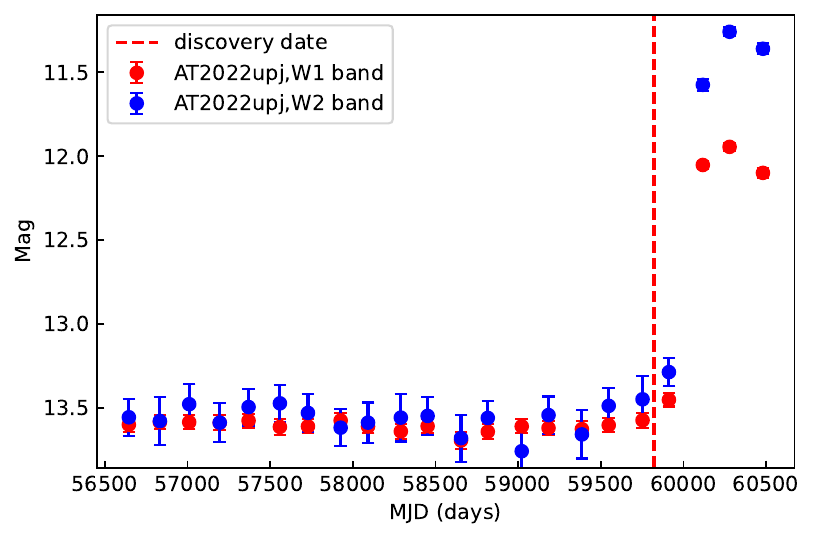}
\includegraphics[width=0.3\textwidth]{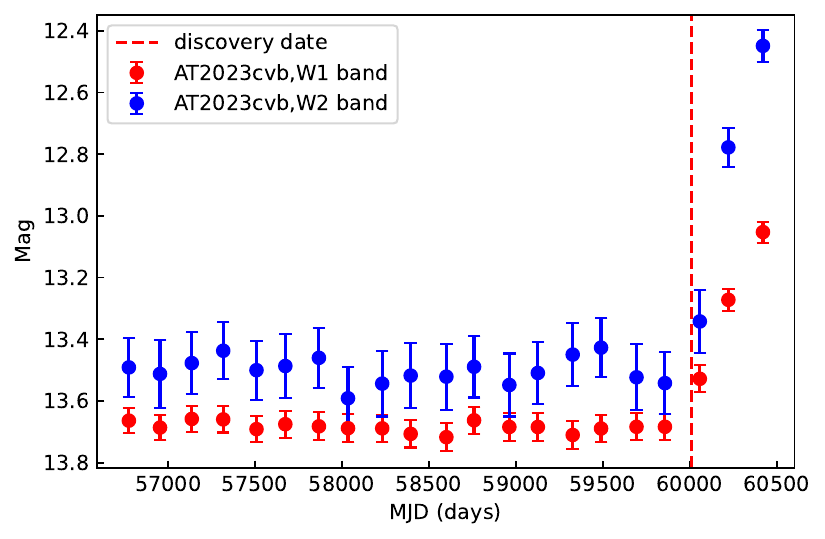}
\caption{The IR light curves of seven ZTF-BTS TDEs with IR echoes, where the red and blue points represent the W1 and W2 band data, respectively. {The red and black vertical lines denote the discovery dates of the TDEs and arrival dates of the associated alert events, respectively.}}
\label{fig:IR light curves}
\end{figure*}

The obtained single-exposure data are initially filtered according to the quality flags indicated in the catalogs. We exclude the bad data points with poor-quality frames ({\tt qi\_fact} $<$ 1), charged particle hits ({\tt saa\_sep} $<$ 5), scattered moonlight ({\tt moon\_masked} $=$ 1), and artifacts ({\tt cc\_flags} $\neq$ 0). Additionally, data is abandoned when multiple PSF components ({\tt nb} $>$ 1 and {\tt na} $>$ 0) are present during photometry, particularly when the source is fitted concurrently with other nearby detections or when a single detection is split into two components during the fitting process. 
To examine the variability of IR emission at the positions of TDEs, we estimate the magnitudes by adopting the median values of the data points. Then we qualify the significance of variability in IR emission at the position of each TDE and take uncertainties into consideration, as given by \citet{2021ApJS..252...32J},
\begin{equation}
\Delta W_{i,\sigma} = (W_{i,\rm max} - W_{i,\rm min})/\sqrt{(W_{i,\rm max,err}^2 + W_{i,\rm min,err}^2) }
\end{equation}
where, $W_{i,\rm max}$ and $W_{i,\rm min}$ are the maximum and minimum magnitude for $i$-th band, respectively. $W_{i,\rm max,err}$ is the error of $W_{i,\rm max}$ and $W_{i,\rm min,err}$ is the error of $ W_{i,\rm min}$.
To identify the IR variability, we impose a requirement that the IR variability, denoted as $\Delta W_{{\rm W1},\sigma}$ or $\Delta W_{{\rm W2},\sigma}$, must exceed 5$\sigma$. We used the following criteria to further select TDEs with IR echoes from these TDEs with IR variability. 1) The IR magnitude presents $>3\sigma$ brightening compared to the median value of the quiescent level before the discovery date \citep{2021ApJS..252...32J}.
2) The brightest magnitude should occur after the discovery dates of TDEs that are obtained from the Transient Name Server (TNS)\footnote{\url{https://www.wis-tns.org/search}}. 
Finally, there are 7 ZTF-BTS TDEs with IR echoes, as listed in Table \ref{tab:IR light curves}. The light curves of these seven TDEs
are presented in Fig \ref{fig:IR light curves}, where red and blue circles are the W1 and W2 band data, respectively. The red dashed lines are the discovery dates of TDEs.

\begin{table*}[!htbp]
	\centering
	\caption{ZTF-BTS TDEs with IR echoes.}
	\begin{tabular}{|c|c|c|c|c|c|c|c|c|c|} 
		\hline
		\text{TNS${\rm_{ID}}$} & \text{Discovery date}$^a$ &\text{MJD}& \text{$\Delta W_{{\rm W1},\sigma}^b$} &
        \text{$\Delta W_{{\rm W2},\sigma}^b$}&\text{$\delta W_1^c$}&\text{$\delta W_2^c$}& $z$&log $\delta L_{W_1}^d$&log $\delta L_{W_2}^d$
        \\
		\hline
        &&(days)&&&(mag)&(mag)&& erg s$^{-1}$&erg s$^{-1}$\\
        \hline        
	AT2019azh& 19/02/22 &58535 & 5.15 & 3.48&0.15 &0.20&0.022&42.04&41.85  \\
        AT2019dsg& 19/04/09 & 58582& 18.59 & 16.40& 0.86   &1.51&0.0512&42.60&42.69  \\
        AT2019qiz& 19/09/19 & 58745 & 38.87 & 41.78  &1.32&2.38&0.0151 &43.21&43.41\\
        AT2020nov& 20/06/27 & 59027& 12.95 & 9.30  &0.56&0.76&0.084&42.31&42.28 \\
        AT2022dyt& 22/02/26 &59635 & 7.96 & 7.37  &0.46&0.79&0.072&42.14&42.18 \\
        AT2022upj& 22/08/31 & 59822 & 30.36 & 16.17  &1.66&2.29&0.054&42.96&42.96 \\
        AT2023cvb& 23/03/06 & 60009 & 7.76 & 6.77  &0.63&1.05&0.071&42.27&42.33 \\
        \hline
	\end{tabular}
\begin{tablenotes}
\item $^a$ The discovery dates of TDEs obtained from TNS.
\item $^b$ The significance of the IR variability between the maximum and minimum magnitudes in the W1 and W2 bands.
\item $^c$ The magnitude difference between the brightest magnitude and the median value of quiescent level.
\item $^d$ The luminosity difference between the brightest magnitude and the median value of quiescent level.
\end{tablenotes}
	\label{tab:IR light curves}
\end{table*}

\subsection{Neutrino sample}
\label{sect:alert sample}
The IceCube Neutrino Observatory has observed high-energy astrophysical neutrinos exceeding the atmospheric background in both cascade and track data.
Since 2016, real-time alerts for individual high-energy events ($>$100 TeV) have been released to the multimessenger observational community through platforms, for example, the General Coordinates Network (GCN)\footnote{\url{https://gcn.nasa.gov/missions/icecube}}.
These alert events concentrate on track-like neutrino candidates, which provide more precise angular localizations compared to cascade events.
Since June 2019, all alert events have been assigned a signalness value and categorized as either gold or bronze based on their probability of astrophysical origin, specifically larger than 50$\%$ and 30$\%$, respectively.
The 90$\%$ containment error radii ($r_{90}$, statistical error only) for these alert events are publicly available. For each alert event, an updated report would be provided after detecting a more accurate position. The latest reported information for each alert event is adopted in our analysis, while we note that one alert event (i.e., IC241016\footnote{\url{https://gcn.nasa.gov/circulars/37794}}) is found to clearly not be of astrophysical origin so we remove it from our sample.
We also note that \citet{IceCube:2023agq} has provided updated $r_{90}$ for neutrinos between 2011 and 2020. However, compared to the catalog covering the period from 2011 to 2020, our primary focus is on the neutrinos published after 2019, as this period suitably matches the ZTF-BTS catalog survey timeline i.e. allowing for alert events occur after the TDEs of ZTF-BTS catalog. Therefore, for our analysis, we adopt the list of all the bronze and gold alert events after 2019 as our sample, leaving at least a 30$\%$ probability of an astrophysical origin for these alert events.

Furthermore, for the association analysis, it is crucial that neutrino events have a good spatial localization. Some studies exclude alert events if their 90$\%$ confidence level error region area, denoted as $\Omega_{90}$, exceeds a threshold value (e.g., \cite{Plavin:2020emb,Hovatta:2020lor}). More recently, rather than directly excluding alert events, \citet{Kouch:2024xtd} proposed a weighting scheme that gradually reduces the effect of alert events with large error regions in the association analysis. 
The weighting scheme removes the need for setting arbitrary selection thresholds and takes into account observed information, including the signalness and $\Omega_{90}$ of alert events in the correlation analysis.

Therefore, in this analysis, we employ two methods to handle our alert events sample. The first method involves noting that AT 2019dsg associates with an alert event of $\Omega_{90}$ $\sim$ 30 deg$^2$ so that we exclude alert events of $\Omega_{90}$ larger than 30 deg$^2$. The other method combines the information of signalness and $\Omega_{90}$ using the weighting scheme to perform this correlation analysis.

\section{Analysis Method}
\label{Sec: Analysis Method}
We perform a systematic search in our samples based on the spatial and temporal criteria. 
The initial step involves examining the spatial correlation between TDEs and alert events. This is realized by cross-matching the coordinates of TDEs with the positions of alert events within their $r_{90}$ containment radius.
In this step, we search for spatially associated events, using the alert events sample with $\Omega_{90} <$ 30 deg$^2$ or the whole alert events sample. According to the $r_{90}$ of IceCube alert events, the calculation of $\Omega_{90}$ for alert events is given by $\Omega_{90} = 41252.96 \times (1 - \cos r_{90})/2$.
In the next step, we identify the temporal coincident events from the spatially correlated events. The TDEs are required to be discovered before the arrival dates of alert events (i.e., $\Delta T>0$, where $\Delta T$ is the arrival date of the alert event minus the discovery date of TDE). 

To test the correlation between neutrino alerts and TDEs, we define a test statistic (TS).
As mentioned in Sec.~\ref{sect:alert sample}, we consider two methods to handle the sample of alert events and calculate the TS value.
The first method involves excluding alert events with $\Omega_{90} >$ 30 deg$^2$ (hereafter referred to CUT method). In this method, the number of TDEs that are spatially-temporally associated with the alert events corresponds to the TS values.
More specifically, if one TDE is found within the $r_{90}$ of an alert event and $\Delta T >$ 0, the TS value would be increased by 1. During the calculation of TS values, we allow a single TDE or alert event to contribute repeatedly to the TS values. Specifically, if one alert event is spatially and temporally associated with $N$ TDEs, or if one TDE is spatially and temporally associated with $N$ alert events, they can each contribute a TS value of $N$ (i.e., the total TS will be plus $N$).

The above CUT method is the simplest approach to determine associations between sources and neutrinos and is widely used \citep{Plavin:2020emb,Hovatta:2020lor,Kovalev:2022izi,Plavin:2022oyy,Suray:2023lsa,Plavin:2023wsb}, but it requires excluding events with large error regions.
To make full use of all neutrino alert events, another method does not apply a $\Omega_{90}$ cut and assigns a weighting factor to each alert (hereafter referred to WEIGHT method). 
We use a similar formalism as in \cite{Kouch:2024xtd} but add an additional Gaussian term to account for the angular distance between the neutrino and the TDE.
The weighting factor is expressed as
\begin{subnumcases}{\cal W_\textit{i} =}
S_i \times \exp(-d_i^2/(2\sigma_i^2))  & \mbox{for} $\Omega_{90,i} \leq M(\Omega_{90})$
\label{eq:case1} \\
S_i \times \exp(-d_i^2/(2\sigma_i^2))  \times \frac{M(\Omega_{90})}{\Omega_{90,i}} & \mbox{for} $\Omega_{90,i} > M(\Omega_{90})$
\label{eq:case2} 
\end{subnumcases}
where $S_i$, $\sigma_{i}$ and $\Omega_{90,i}$ are the signalness, $1\sigma$ containment radius estimated from $r_{90}$ and $\Omega_{90}$ for the $i$-th alert event, respectively;
$d_i$ is the angular distance between TDE and central position of alert event;
$ M(\Omega_{90})$ is the median of $\Omega_{90}$ of alert events sample.
In Eq.~(2), the term $S_i$ accounts for the probability that the alert is an astrophysical neutrino rather than a background event.  
The term $\exp(-d_i^2/(2\sigma_i^2))$ considers the location of the TDE within the confidence region of a correlated neutrino; namely, associations that are closer in distance are given higher weights.  
The term ${M(\Omega_{90})}/{\Omega_{90,i}}$ takes into account the fact that an event with a larger error region indicates lower data quality and a higher chance of random association, and thus should be assigned a lower weight.
In the WEIGHT method, the TS value is calculated by ${\rm TS}=\sum_i {\cal W}_i$ with ${\cal W}_i$ given by Eq.~(2) when TDEs are found to be spatially-temporally associated with the $i$-th alert event.
More specifically, the TS value would be adjusted from 1 in the CUT method to $\cal W_\textit{i}$ for each TDE associated with the $i$-th alert event in order to reduce the effect of alert events with large $\Omega_{90}$ in the WEIGHT method.

\begin{figure*}
\center
\includegraphics[width=0.6\textwidth]{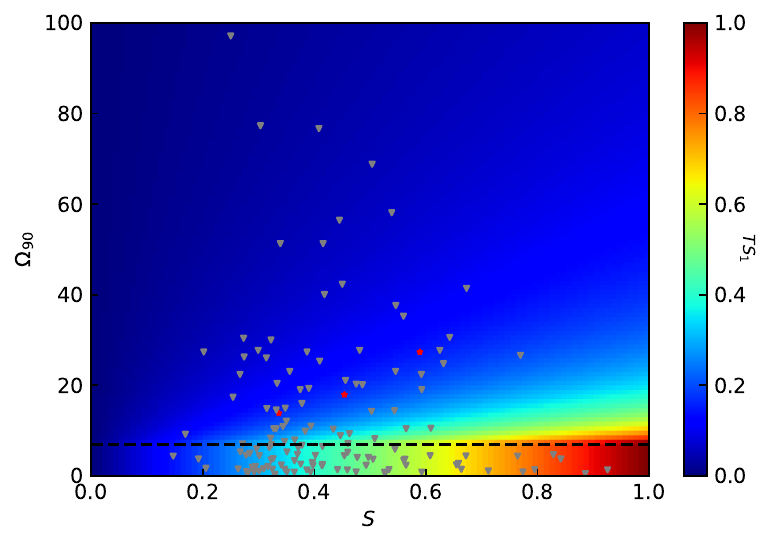}
\caption{This plot shows the distribution of the IceCube alert events in the $\Omega_{90}-S$ plane, where $\Omega_{90}$ is the event's error region size and $S$ is the signalness.
The red stars represent the alert events that are spatially associated with at least one TDE (i.e., the TDE is within the error region of the event), while the inverted triangles are alert events that are not associated with any TDE.
The black dashed line indicates the median $\Omega_{90}$ value ($6.88\,{\rm deg^2}$) of our alert events sample. This plot only includes alert events where $\Omega_{90} <$ 100 deg$^2$.
The color map shows the 1-event TS value ($TS_1$) that an alert event could contribute to the total TS value (the angular distance term is set to 1)} in the WEIGHT method of the correlation analysis,
which depends on the $\Omega_{90}$ and $S$ of the alert event.
\label{fig:heating_map1}
\end{figure*}

According to the $r_{90}$ of the alert events, we estimate that the $ M(\Omega_{90})$ of our alert events sample is 6.88 deg$^2$. We present the 1-event TS values using the heating map after considering the weight factor (the angular distance term is not included because it depends not only on the event properties) in Fig.~\ref{fig:heating_map1}, where the color bar is the TS value when one TDE is found to be associated with an alert event, the black dashed line indicates the median value of our alert events sample (6.88 deg$^2$), inverted triangles are alert events that are not associated any TDE and the red stars present alert events that are spatially-temporally associated with at least one TDE. One can see that the $\Omega_{90}$ values of the alert events that are associated with TDEs are larger than the median value.

Based on these two methods for handling the alert sample and calculating the TS values, we perform Monte Carlo simulations to quantify the correlation between the TDE population and the alert population. 
Since in our previous work we found that different randomization methods in the MC simulations may have some effect on the results of the correlation analysis \citep{Lu:2024flp}, in the following analysis we will test three different randomization methods.

Method I: Similar to previous works \citep[e.g.,][]{Plavin:2020emb,Hovatta:2020lor}, due to the sensitivity of IceCube only depending on the zenith angle \citep{IceCube:2016tpw}, we only randomize the RA of the alert events while keeping the DEC, $r_{90}$ and arrival date fixed to their observed values, and the coordinates of the TDEs also keep unchanged.

Method II: We not only randomize the RA of alert events, as in Method I, but also randomize the positions of the TDEs using the method described in \citet{Lu:2024flp}. 
For each mock TDE source, we randomly select an $\rm RA'$ and a $\rm DEC'$ from the observed lists of RA and DEC, then impose an additional random shift to $\rm (RA', DEC')$ within a region of $10^\circ$ radius to obtain the position of the mock source. In the simulations, we require the simulated sources not to be outside the sky region of the ZTF-BTS sample, which covers the northern sky with a DEC range from $-30^\circ$ to 90$^\circ$ but excluding the Galactic plane region ($|b| < 7^\circ$). We also require the number of mock sources to be the same as the real sample. 
Finally, we perform KS tests on $\rm RA'$ and $\rm DEC'$ distributions to ensure that the mock sources preserve the same distribution pattern of the actual sources.

Method III: The same as Methods I and II, we randomize the RA of alert events. For TDEs, the method used to simulate the mock sample is the same as \citet{Buson:2022fyf, Buson:2023irp, Bellenghi:2023yza}. We generate mock TDEs by randomizing the positions of the sources within a 10$^\circ$-radius range from their original positions. During this procedure, we also require that each mock source remains within the sky region of the ZTF-BTS sample. The KS tests are performed in the same manner as in Method II.

By performing 10$^4$ MC simulations, we obtain the TS distribution. We can then estimate the chance probability (i.e., the $p$ value of rejecting the null hypothesis) according to the fraction of TS values that are equal to or larger than the observed one,
\begin{equation}
p = \frac{M+1}{N+1}
\end{equation}
where $M$ is the number of TS values equal to or larger than the observed one and $N$ is the total number of MC simulations \citep{Davison_Hinkley_1997}.

\section{Results}
\subsection{Evaluate the correlation between ZTF TDEs and neutrino alerts}
We first systematically search for associated events that satisfy spatial and temporal criteria.
After adopting the CUT method, there are 122 alert events remaining in our sample. Two associated events are found, and all of these events have IR echoes, which are listed in Table \ref{tab:tab1}. 
{Apart from AT2019dsg, another TDE (AT2019azh) is also found to be associated with a neutrino alert (IC230217A).
The association between AT2019azh and IC230217A has not been reported before and is newly discovered in this work. 
We will further discuss their connection in Sec.~\ref{sec:at2019azh}.}

Under the WEIGHT method, all 143 alert events are utilized to search for associations. We find that there are six occurrences of TDEs within the $r_{90}$ of five alert events. Among them, one TDE (AT2019azh) is located within $r_{90}$ of two alert events. Additionally, the $r_{90}$ of two alert events each includes two TDEs. However, we note that the $r_{90}$ values of these two alert events are significantly large ($>$ 10$^\circ$), contributing only negligible TS values. In total, five distinct TDEs are found within $r_{90}$ of four alert events, two of which have IR echoes, as listed in Table \ref{tab:tab2}.
Apart from AT2019dsg, the other TDE that contributes a relatively large TS in the WEIGHT method is still AT2019azh.

\begin{table*}[!htbp]
	\centering
	\caption{Associated events when excluding alert events with error regions larger than 30 deg$^2$.}
	\begin{tabular}{|c|c|c|c|c|c|c|c|c|c|c|}
	\hline
Alert & \text{RA$_{\rm alert}$} & \text{DEC$_{\rm alert}$} & \text{$r_{90}$}&
    \text{$T_{\rm alert}$} & Signalness & \text{TDE event} &\text{$\Delta T^{a}$}& \text{TS$_1$ value$^b$} &\text{IR echoes$^c$}&$z$\\
    \hline
    &(deg) & (deg)&(deg)& &&&(days)&&&
    \\
	\hline
IC230217A & 124.54& 20.74 & 2.39 & 23/02/17 & 0.454 & AT2019azh & 1456 &1 &Yes&0.022  \\
IC191001A & 314.08& 12.94 & 2.95 & 19/10/01 & 0.589 & AT2019dsg  & 175 &1&Yes&0.0512  \\
    \hline
	\end{tabular}
\begin{tablenotes}
\item $^a$ The time delay between the alert event arrival date $T_{\rm alert}$ and the TDE discovery date.
\item $^b$ In the CUT method, each association contributes $TS=1$ to the total TS value of the correlation. 
\item $^c$ This column indicates whether the TDEs have IR echoes.
\end{tablenotes}
	\label{tab:tab1}
\end{table*}

\begin{table*}[!htbp]
	\centering
	\caption{Associated events considering the whole alert sample in the WEIGHT method.}
	\begin{tabular}{|c|c|c|c|c|c|c|c|c|c|c|c|}
		\hline
Alert & \text{RA$_{\rm alert}$} & \text{DEC$_{\rm alert}$} & \text{$r_{90}$} & \text{Date$_{\rm alert}$} & \text{Signalness} & \text{TDE event} &\text{$\Delta T^{a}$}& \text{$d^{b}$}&\text{TS$_1$ value$^c$}& \text{IR echoes$^e$}&$z$\\
\hline
&(deg) & (deg)&(deg)&(yy/mm/dd)&&&(days)&(deg)&&& \\
\hline
IC240307A&239.63& 39.94 & 15.38 & 24/03/07& 0.606  & AT2020vwl& 1244 &14.19& 7.90$\times 10^{-4}$&No&0.0325 \\
IC240307A&239.63& 39.94 & 15.38 & 24/03/07& 0.606  & AT2020pj& 1526 &8.72&  2.68$\times 10^{-3}$&No&0.068 \\
IC230707B&127.18& 20.74 & 10.16 & 23/07/07& 0.466  & AT2022bdw & 522 &2.31&  8.78$\times 10^{-3}$&No&0.03782 \\
IC230707B&127.18& 20.74 & 10.16 & 23/07/07& 0.466  & AT2019azh & 1596 &4.06&  6.85$\times 10^{-3}$&Yes&0.022 \\   
IC230217A&124.54& 20.74 & 2.39 & 23/02/17 & 0.454 & AT2019azh & 1456 &2.22&  2.39$\times 10^{-2}$ &Yes&0.022  \\
IC191001A&314.08& 12.94 & 2.95 & 19/10/01 & 0.589 & AT2019dsg  & 175  &1.28& 9.63$\times 10^{-2}$&Yes&0.0512 \\
        \hline
	\end{tabular}
\begin{tablenotes}
\item $^a$ The time delay between the alert event arrival date $T_{\rm alert}$ and the TDE discovery date.
\item $^b$ The angular distance between alert event and TDE.
\item $^c$ The 1-event TS represents the value that an association event can contribute to the total TS of the correlation.
\item $^e$ This column indicates whether the TDEs have IR echoes.
\end{tablenotes}
	\label{tab:tab2}
\end{table*}

\begin{figure*}
\center
\includegraphics[width=0.45\textwidth]{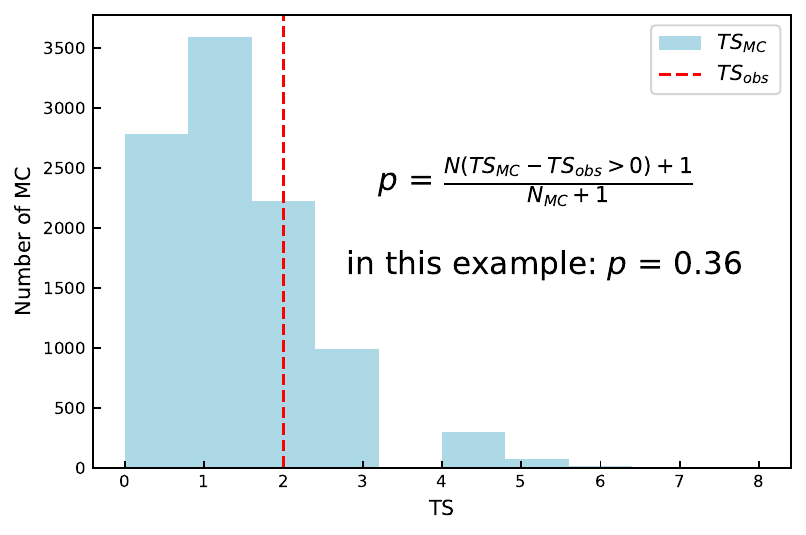}
\includegraphics[width=0.45\textwidth]{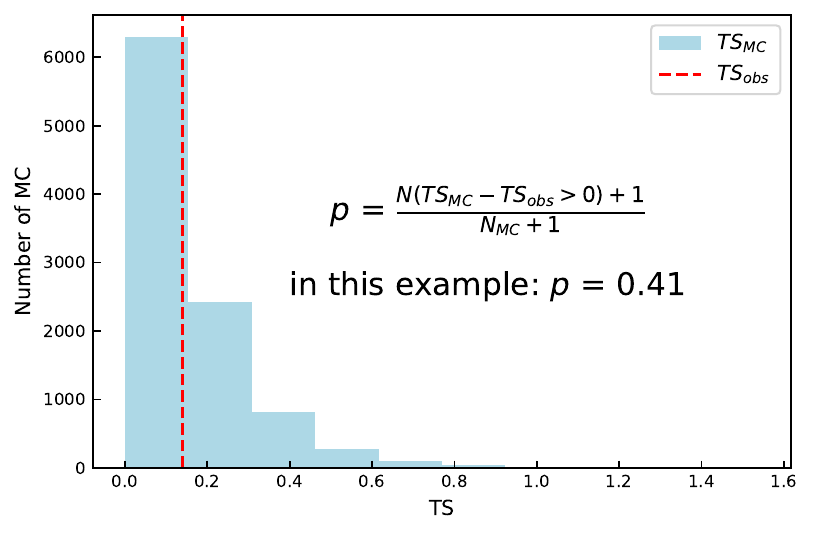}
\caption{The example (MC method III+All TDEs) for TS distribution obtained from 10$^4$ MC simulations, where the left panel and right panel are the TS distribution using CUT and WEIGHT methods, respectively.}
\label{fig:TS_distribution}
\end{figure*}

We employ three MC simulation methods, as described in Sec. \ref{Sec: Analysis Method}, to estimate the $p$ values for these search results. 
Each estimation is performed using 10$^4$ times MC simulations. For instance, if using the all TDEs sample and simulating Method III, the TS distribution obtained from 10$^4$ times MC simulations is shown in Fig \ref{fig:TS_distribution}, where the left panel illustrates the TS distribution when the CUT method is applied, and the right panel corresponds to the WEIGHT method. The blue region represents the simulated TS distribution, while the red dashed line indicates the observed TS value. The region to the right of the red dashed line corresponds to simulated TS values that are equal to or greater than the observed TS value.

\begin{table*}[!htbp]
	\centering
	\caption{The $p$-values of the TDE-neutrino correlation obtained in different MC methods.}
	\begin{tabular}{|c|c|c|c|c|}
		\hline
		& \text{Method I} & \text{Method II} & \text{Method III} & \text{TDE sample$^a$}\\
		\hline
	       CUT method&0.38& 0.36 & 0.36  &All TDEs  \\
          WEIGHT method &0.39& 0.42 & 0.41 &All TDEs\\
          CUT method&0.026& 0.014 & 0.018  &TDEs with IR echoes \\
          WEIGHT method &0.039& 0.042 & 0.052 & TDEs with IR echoes\\
         
        \hline
	\end{tabular}
\label{tab:tab3}
\end{table*}
By realizing 10$^4$ times MC simulations, the results obtained from the three MC methods are summarized in Table \ref{tab:tab3}, which includes the results from both the CUT method and the WEIGHT method. Using the all TDEs sample, the best $p$-values are 0.36 and 0.39 for the CUT method and WEIGHT method, respectively. When using the subsample of TDEs with IR echoes, the best $p$-values are 0.014 and 0.039 for the CUT method and WEIGHT method, corresponding to 2.46$\sigma$ and 2.06$\sigma$, respectively. The most significant correlation with neutrinos is obtained from the subsample of TDEs with IR echoes at a $\sim 2.46\sigma$ confidence level, which may indicate that TDEs with IR echoes are more likely to correlate with IceCube neutrinos compared to the all-TDE sample.

\subsection{The systematic errors of alert events}
\label{sec:sys}
Although the IceCube collaboration strives to consider all potential errors in estimating the size of the high-energy neutrino error regions, there may still be some unknown systematic errors, such as those arising from ice inhomogeneities, that cannot be completely addressed \citep[e.g.][]{IceCube:2023htm}. An upper limit of 1$^\circ$ is estimated for this systematic error by \citet{IceCube:2013cdw}. While efforts to constrain this parameter are ongoing \citep{Abbasi:2021ryj,IceCube:2022der}, the most recent findings suggest that the associated error is likely negligible \citep{IceCube:2022der,IceCube:2023htm}.

In this work, we add a systematic error upper limit of 1$^\circ$ in quadrature to the four directions of RA and DEC, so that the final uncertainty in each direction is $\sqrt{{r_{90}}^2 + (1^\circ)^2}$. We then search for associated events among all alert events (having added 1$^\circ$ systematic error) and TDEs. However, no additional associated events are found when compared to those listed in Table \ref{tab:tab2}. 
Therefore, the $p$ value is expected to increase as the systematic error grows larger within the range of $0^\circ \sim 1^\circ$.
When a 1$^\circ$ systematic error is added to each alert event, the $ M(\Omega_{90})$ of the alert events sample is 10.02 deg$^2$. We only adopt the WEIGHT method to test the effect of systematic error on the estimation of $p$ values.
The $p$ values estimated by three MC simulation methods are listed in Table \ref{tab:tab5}, showing an increase of $\sim 7\%$ to $\sim 37\%$ compared to those in Table \ref{tab:tab3}.

\begin{table*}[!htbp]
	\centering
	\caption{$p$-value results when introducing 1$^\circ$ systematic error or time delay factor.}
	\begin{tabular}{|c|c|c|c|c|}
		\hline
		&\text{Method I} & \text{Method II} & \text{Method III}& \text{TDE sample$^a$}\\
		\hline
		 WEIGHT+Syst.&0.48& 0.50 &0.49  &All TDEs  \\
       WEIGHT+Syst.& 0.046 & 0.057  & 0.056  & TDEs with IR echoes  \\
         WEIGHT+Delay &0.22& 0.22 & 0.20 &All TDEs  \\
         WEIGHT+Delay &0.0077& 0.027 & 0.020 &TDEs with IR echoes \\
        \hline
	\end{tabular}
	\label{tab:tab5}
\end{table*}

\subsection{Time delay factor}
In the above analysis, the temporal coincidence criterion only requires that the discovery dates of TDEs are earlier than the arrival dates of alert events. Under this criterion, we estimate the $p$ values of correlation between TDEs/subsample of ZTF-BTS catalog and IceCube neutrinos, and the most significant correlation is given by the subsample of TDEs with IR echoes.

However, observationally, the time delays of potentially associated events between TDEs/TDE candidates and IceCube neutrinos are reported to range from several months to years. For instance, one bona fide TDE, AT2019dsg shows a time delay of $\sim$ half a year. The other two TDE candidates i.e. AT2019fdr and AT2019aalc show a time delay of $\sim$ half a year and $\sim$ one year after the discovery dates of TDE candidates, respectively. AT2021lwx also presents a time delay of $\sim$ one year. Although the accurate time delay still remains obscure due to the limited number of potentially associated events reported so far, these events present a time delay within a span of $\sim$ one year.

Therefore, we tentatively impose a time delay factor into the weighting scheme to reduce the weight of associated events with large time delays. By combining an additional time delay factor into the weight factor of Eq. \ref{eq:case1}, \ref{eq:case2}, we reduce the impact of events with large time delays, i.e.
\begin{subnumcases}{\cal W_\textit{i} =}
S_i/\Delta T \times \exp(-d_i^2/(2\sigma_i^2))  & \mbox{for} $\Omega_{90,i} \leq M(\Omega_{90})$ \\
S_i/\Delta T \times \exp(-d_i^2/(2\sigma_i^2))  \times \frac{M(\Omega_{90})}{\Omega_{90,i}} & \mbox{for} $\Omega_{90,i} > M(\Omega_{90})$
\label{eq:eq4}
\end{subnumcases}

After adding the time delay factor, the TS values are mainly contributed by the AT2019dsg event.
By realizing 10$^4$ times MC simulations, the $p$ values for the three MC simulation methods are summarized in Table \ref{tab:tab5}.
Using the WEIGHT method, compared to the results without the time delay factor, the best $p$ value of the all-TDEs sample decreases to 0.20, while that for the TDEs with IR echoes subsample decreases to 0.0077 ($\sim 2.66\sigma$). After introducing the time delay factor, the subsample with IR echoes also presents the most significant correlation with IceCube neutrinos compared to the all-TDEs sample.

\section{Discussion}
\label{sec:disscussion}

{In Table \ref{tab:tab6}, we summarize the properties of TDEs associated with alert events found in previous works and in this work. We find that most of the listed properties are consistent for all the TDEs.
For these TDEs, all exhibit bright IR echoes, and except for the two TDE candidates identified by \citet{Jiang:2023kbb} in the IR data, all have been reported in both optical and X-ray observations.
Except for AT2019azh, the time differences between IR peaks and arrival dates of alert events range from $\sim$18 to 159 days in the source frame. The arrival dates of alert events are closer to the IR peaks than to the optical peaks ($\sim$148 to 228 days in the source frame except AT2019azh), supporting the connection between IR photons and neutrinos. 
}

\subsection{On the connection between high-energy neutrinos and IR emissions}
\label{sec:nu_ir}
TDEs have increasingly been discovered through optical transient surveys and have been proposed as potential sources of high-energy neutrinos over the past few years.
Interestingly, it is noted that the first three TDE candidates reported to be possibly associated with high-energy neutrinos, including the confirmed TDE AT2019dsg and the candidates AT2019fdr and AT2019aalc \citep{2021NatAs...5..510S, Reusch:2021ztx, vanVelzen:2021zsm}, as well as some other similar events reported subsequently \citep{Jiang:2023kbb,Yuan:2024foi,Li:2024qcp}, all have IR echo emission being detected. 
Here we discuss the possible connection between high-energy neutrinos and IR echo emissions.
In this work, our analysis also shows that the TDEs with IR flares display a stronger correlation with high-energy neutrinos, which further supports previous findings that the TDEs coinciding with neutrinos tend to have prominent IR echo emissions.

The IR echo emission typically originates from a dust torus surrounding the supermassive black hole: the optical-ultraviolet or X-ray radiation from the TDE outburst is absorbed by the dust and re-emitted in the IR band, indicating the presence of a dense environment around such TDEs \citep{Lu:2015jfg,vanVelzen:2016jsk,Jiang:2016znw}.
Although not fully clear, possible reasons for TDEs with IR echoes are more likely to be associated with IceCube high-energy neutrinos have been widely discussed \citep{Wu:2021vaw,Mou:2020csb,Winter:2022fpf,Yuan:2023cmd,Yuan:2024foi}:
(1) Significant IR echoes imply a dusty/cloudy environment around the black hole, where winds or outflows interacting with the circumnuclear medium can drive the formation of shocks, which accelerate protons to high energies and lead to neutrino production.  
(2) The dense gas and dust in the dust torus themself provide target material for the accelerated high-energy protons. Compared to TDEs without a dust torus, such an environment significantly enhances the rate of neutrino production via $pp$ interactions.  
(3) The IR photons produced by dust echoes serve as target photons for $p\gamma$ interactions, and accelerated protons interact with the IR photons, though this requires the protons to be accelerated to sufficiently high energies.

Overall, our analysis, together with the previous works, suggests that the IR echoes are not only an important feature of TDEs, but may also reflect key environmental conditions for neutrino production. Searching for neutrino emission from transients with infrared echoes is therefore a promising direction worth further attention.
\begin{table*}[!htbp]
	\centering
	\caption{A summary of the properties of TDEs associated with neutrino alerts.}
	\begin{tabular}{|c|c|c|c|c|c|c|c|c|c|}
		\hline
		\text{TDE event} &\text{$\Delta T_{\rm OP,peak}^a$}&\text{$\Delta T_{\rm IR,peak}^b$}& \text{log $L_{\rm IR,peak}^c$}&log $L_{\rm X,peak}^d$ &\text{$z$}  &\text{log $M_{\rm{BH}}$} &\text{$E\rm_{alert}$} &Ref \\
		\hline
        &(days) & (days)&(erg s$^{-1}$)&(erg s$^{-1}$)&&(M$_{\odot}$)&(TeV)& \\
        \hline
		AT2019dsg&150& -19 & 43.30 &43.4& 0.0512&6.74& 217.42&(5),(6) \\
        AT2019fdr& 289 & -75  & 44.56  &43.1& 0.267& 7.10& 82.186&(4),(6)\\
        AT2019aalc &148& -91 & 43.80 &42.1&0.0360&7.23&176.48& (6)\\
        AT2021lwx &370& 317 & 45.49 & 45.2& 0.995&8.00&106.32& (7),(8)\\
        SDSSJ1048+1228 &-& -120 & 42.84 &-&0.0537&7.86&375.23&(2)\\
        SDSSJ1649+2625 &-& -69 & 43.20 &-&0.0588&6.49&82.186 &(2)\\
        AT2019azh &1556& 1560 & 42.06 &43.1&0.0223&6.50&154.28 & This work, (1),(2)\\
        AT2019azh &1416& 1420 & 42.06 &43.1&0.0223&6.50&54.85 & This work, (1),(2)\\
        \hline
	\end{tabular}
\begin{tablenotes}
\item 
	   The references are as follows: 
        (1)\citet{Hinkle:2020ilv},
          (2) \citet{Jiang:2023kbb},
          (3)\citet{Liu:2019dnt}
	   (4) \citet{Reusch:2021ztx}, 
        (5) \citet{2021NatAs...5..510S},
          (6) \citet{vanVelzen:2021zsm},
	   (7) \citet{Wiseman:2023tnv}, and
       (8)\citet{Yuan:2024foi}.
\item $^a$ The arrival date of alert event subtracts the date of TDE optical peak.
\item $^b$ The arrival date of alert event subtracts the date of TDE IR peak.
\item $^c$ The luminosity values are the total of W1 and W2 bands.
\item $^d$ The X-ray luminosity covers an energy range of 0.3-10 keV for AT2021lwx and AT2019azh, while the others have an energy range of 0.2-10 keV.
\end{tablenotes}
	\label{tab:tab6}
\end{table*}

\subsection{The association between AT2019azh and IC230217A}
\label{sec:at2019azh}
In this work, we find that AT2019azh is spatially associated with two alert events (IC230217A and IC230707B). 
Among the two, IC230707B has a very large localization uncertainty (10.16 degree radius, corresponding to an area of $\sim$323 deg$^2$), which will result in a high probability of coincidental association with TDEs.
As shown in Table~\ref{tab:tab2}, there are two TDEs (AT2019azh and AT2022bdw) associated with IC230707B within its location error circle, and AT2022bdw is even closer to the nominal position of IC230707B and has a shorter time difference $\Delta T$. Considering these, individually discussing such events like IC230707B with poor localization is not particularly meaningful.
Therefore, we focus on discussing the association between AT2019azh and the alert IC230217A.

{AT2019azh is a thermal TDE exhibiting unusually long-lived radio emission and delayed X-ray brightening \citep{Goodwin:2022zik}. Its optical light curve peaked $\sim$27 days after discovery.  
Radio observations revealed a gradual brightening over $\gtrsim2$ years post-disruption, followed by a decline attributed to either interaction with the circumnuclear medium or the shutdown of the central engine \citep{Goodwin:2022zik}. Even 850 days after the TDE, AT2019azh still exhibited strong radio emission. The inferred radio outflow is non-relativistic, plausibly a quasi-spherical or mildly collimated wind from accretion. 
These radio observations confirm that particle acceleration is occurring. 
AT2019azh shares many similar characteristics with AT2019dsg \citep{Stein:2020xhk}: both are thermal TDEs with sub-relativistic outflows, and their parameters in Table~\ref{tab:2019azh} show significant similarity. 
AT2019azh also displays a rare bright X-ray emission among the TDE population, with a peak luminosity of $\sim $1.2$\times 10^{43}$ erg/s \citep{Liu:2019dnt,Hinkle:2020ilv}, similar to those observed in AT2019dsg, AT2019fdr, and AT2019aalc. 
The late-time radio behavior of AT2019azh is also similar to that of AT2019dsg \citep{Stein:2020xhk}.
Both TDEs show initially increasing energy and flux density attributed to constant energy injection from the outflow.
These similarities suggest that AT2019azh is also a potential neutrino-emitting source.}

{However, compared to AT2019dsg, the neutrino event associated with AT2019azh occurred $\sim$1500 days after the onset of the TDE, and such a long time delay challenges the possibility that the neutrino originated from AT2019azh. 
Considering that the IR echo emission from this TDE was only observed at relatively early stages (see Fig.~\ref{fig:IR light curves}), if AT2019azh is indeed associated with IC230217A, then its neutrino production might differ from that of AT2019dsg, AT2019fdr, AT2019aalc \citep{Winter:2022fpf} and AT2019lwx \citep{Yuan:2024foi} for which, as discussed in Sec.~\ref{sec:nu_ir}, IR photons may serve as the target photons in the neutrino production process.}

{Although there is a significant delay between IC230217A and the TDE, the radio and X-ray emissions from AT2019azh last for a very long time, making neutrino production still plausible at $\sim 1500\,{\rm days}$. 
The radio signal typically serves as a tracer for the particle acceleration processes required for neutrino generation. 
The neutrino energy of IC230217A reported by IceCube is 54.9 TeV.
This corresponds to parent proton energy of $\sim$1 PeV under the typical assumption of a conversion fraction of $\eta_{p\nu}\sim0.05$ from parent protons to neutrinos. 
Based on the velocity and size of the outflow given by the final observational epoch in Table 2 of \citep{Goodwin:2022zik}, the Hillas criterion $E_{\max} \sim Z e B R \beta$ indicates that the outflow is capable of accelerating protons to $>\rm PeV$.
Additionally, to produce neutrinos with an energy of 54.9 TeV, the required target photon energy is $\epsilon_{\gamma} \sim ({m_{\Delta}^2 - m_p^2})/{4 \epsilon_p} \sim 3\,{\rm keV}$ and is in the X-ray band.
The detection of X-ray emission even at $\sim$1500 days post-disruption \citep{2024ApJ...966..160G}, albeit at low flux, suggests that neutrino production via $p\gamma$ interactions remains possible. 
However, the neutrino yield from late-time $p\gamma$ interactions may be relatively low. Another possible neutrino production mechanism is $pp$ interactions, where protons in the circumnuclear medium are more likely to serve as target particles in the hadronic processes.
For such a channel, interactions between the outflow and surrounding materials (e.g., unbound stellar debris or clouds) form shocks and lead to particle acceleration. The accelerated protons could then interact with also these materials and produce neutrinos \citep{Murase:2020lnu,Wu:2021vaw,Winter:2022fpf}.
Neutrinos produced via the $pp$ process typically have lower energies, which is also consistent with the case of IC230217A.}

{In all, the optical, radio, and X-ray emissions detected at a late stage suggest activity of the central engine or interactions between the outflow and the circumnuclear medium at around 1500 days, which may generate shocks and accelerate protons, leading to neutrino production. Therefore, despite the large time delay, AT2019azh remains a promising neutrino source candidate.}

\begin{table}[h]
\centering
\caption{The properties of AT2019azh and AT2019dsg.}
\begin{tabular}{l c c l}
\toprule
 & \textbf{AT2019azh} & \textbf{AT2019dsg} & \textbf{References} \\
\midrule
$L_{\rm radio}$ [erg/s] & $10^{38}$ & $2\times10^{38}$ & (1), (1) \\
$L_{\rm X}$ [erg/s] & $10^{43}$ & $2.8\times10^{43}$ & (1), (3) \\
$L_{\rm blackbody}^a$ [erg/s] & $10^{44.3}$ & $10^{44.0}$ & (2), (2) \\
$R_{\rm blackbody}^b$ [cm] & $10^{14.9}$ & $10^{14.7}$ & (2), (2) \\
$T_{\rm blackbody}^c [K]$ & $2.7\times10^{4}$ & $2.8\times10^{4}$ & (2), (2) \\
$E_{\rm out}^d$ [erg] & $10^{49}\ {\rm to}\ 10^{51}$ & $2.5\times10^{49}\ {\rm to}\ 2\times10^{50}$ & (1), (3) \\
$\beta_{\rm out}^f$ & $0.1-0.2$ & $0.1-0.2$ & (1), (3) \\
$B^g$ [G] & $\sim 0.01-0.1$ & $\sim 0.07$ & (1), (3) \\
$R_{\rm out}^h$ [cm] & $10^{17}-3\times10^{17}$ & $7\times10^{16}$ & (1), (3) \\
$n_{\rm e}^i$ [cm$^{-3}$] & $10^{2}-3\times10^{3}$ & $10^{3.4}$ & (1), (1) \\
Redshift & 0.0223 & 0.0512 & (3), (3) \\
\bottomrule
\end{tabular}
\begin{tablenotes}
\item 
	   The references are as follows: 
        (1) \citet{Goodwin:2022zik},
        (2) \citet{Lin:2022jvw},
        (3) \citet{Stein:2020xhk}.
        See also:
        (4) \citet{Hinkle:2020ilv},
        (5) \citet{vanVelzen:2020hrd}.
\item $^{a \sim c}$ Blackbody luminosity, radius and temperature.
\item $^d$ The energy of the outflow, estimated using radio observations during the period from 132 to 850 days post-disruption for AT2019azh and from 42 to 178 days for AT2019dsg.
\item $^f$ The same as $^d$, but for the outflow velocity in units of $c$.
\item $^g$ For AT2019azh, this row presents the estimated magnetic field during the 132-850 day period post-TDE, while for AT2019dsg it corresponds to the estimation at the arrival date (178 day) of the neutrino.
\item $^h$ The same as $^g$, but for the derived radius of the outflow.
\item $^i$ The same as $^g$, but for the ambient electron density.
\end{tablenotes}
\label{tab:2019azh}
\end{table}

\section{Summary}
\label{sec:summary}
Currently, the main contributors to the IceCube neutrino flux are still obscure. Among various astrophysical sources, TDEs have been suggested as possible emitters of IceCube neutrinos. Therefore, investigating the correlation between the TDE population and IceCube neutrinos could help us better understand whether TDEs could really serve as neutrino emitters. In this work, we study the correlation between the TDEs/TDE candidates from the ZTF-BTS catalog and the population of IceCube neutrinos. 
Considering that good spatial localization is crucial for establishing a reliable association, we employ two methods to handle the alert events with large error circles. One method involves directly excluding alert events with error area $\Omega_{90} >$ 30 deg$^2$, while another method introduces a weighting factor. 
We quantify the significance of the correlation using MC simulations. 
We find that the subsample of TDEs with IR echoes presents the most significant correlation with IceCube neutrinos, though still below the $3 \sigma$ confidence level. We also discuss the case that includes a systematic error in the events' error radii. Given that only the statistical error is publicly available for alert events, we add a systematic error of $1^\circ$, which is the upper limit of the systematic error estimated by \citet{IceCube:2013cdw}, in quadrature to the statistical error. There are no additional associated events arising from the inclusion of $1^{\circ}$ systematic error and the obtained results could be treated as the upper limits of the $p$ values. 
Furthermore, for the potentially associated events reported so far, the arrival dates of neutrinos are within an interval of $\sim1$ year after the discovery dates of the TDEs/TDE candidates. Therefore, we tentatively introduce a time delay ($\Delta T$) term into the weighting factor in our MC simulation to reduce the weight of large time-delayed events. By combining the time delay factor, the best $p$ values decrease to 0.20 and 0.0077 ($\sim 2.66\sigma$) for the all-TDE sample and the subsample with IR echoes, respectively. However, the lowest $p$ value remains below the 3$\sigma$ confidence level, even though trial factors have not been taken into account yet. 

In total, at present, no significant correlation is found in the population analysis between the ZTF-BTS TDEs and IceCube alert neutrino events. Our work is the first research to quantify the $p$-value of the TDE-neutrino association with actual TDE and neutrino observations.
We find that the TDEs with IR flares have a stronger correlation with high-energy neutrinos, consistent with previous findings.
Another finding of this work is the discovery of a new TDE–neutrino association: AT2019azh and IC230217A.
Although IC230217A occurred a relatively long time after the IR/optical peak of AT2019azh, the TDE exhibits unusually long-lived radio and X-ray emission, making it still plausible that IC230217A originated from AT2019azh.
This represents another neutrino alert discovered to be associated with a TDE.

In our sample, only seven TDEs have IR flares and two of them satisfy the spatial-temporal association with high-energy neutrinos. The number of TDEs with IR flares is still limited, which hampers us from obtaining conclusive results.
In the future, the persistent optical survey conducted by ZTF-BTS \citep{Perley:2020ajb} is expected to detect more TDEs, in the order of 10 TDEs per year. 
Moreover, hundreds to thousands of TDEs are expected to be discovered per year by advanced surveys such as the Legacy Survey of Space and Time (LSST) at the Vera Rubin Observatory \citep{2019ApJ...873..111I,2020ApJ...890...73B} and the Wide Field Survey Telescope (WFST, \cite{2022MNRAS.513.2422L}; \cite{2023SCPMA..6609512W}). 
It is also expected that the planned NEO Surveyor \citep{2023PSJ.....4..224M}, the successor to the NEOWISE mission, will detect more IR flares and help build a larger sample of TDEs with IR echoes, addressing the current issue of a too small sample size.
{Furthermore, neutrino events with better signalness and directional resolution are also crucial to solve the TDE-neutrino correlation mystery. The next-generation neutrino observatories would also play important roles, such as IceCube-Gen2 \citep{IceCube:2014gqr}, KM3NeT \citep{KM3Net:2016zxf}, GRAND \citep{Fang:2017mhl}, Baikal-GVD \citep{Baikal-GVD:2018isr}, and RNO-G \citep{RNO-G:2020rmc}.}
By identifying more cases of associations between IR-TDEs and high-energy neutrinos, we will hopefully further uncover the intrinsic connection between the TDE population and high-energy neutrinos, and understand the role that IR echoes play in the neutrino emission process.

{{\it On the very day we were about to submit our reply to the second-round review, we noticed that an IceCube Collaboration’s proceeding for ICRC 2025 appeared online \citep{zegarelli2025icecat2updatedicecubeevent}. This proceeding pointed out that the updated version of the alert data has smaller error circles, which places AT2019dsg, AT2019fdr and AT2019aalc outside the error circles, and therefore no longer associated with the neutrino events. This challenges the possibility that these sources are potential neutrino emitters. However, since the IceCat-2 data has not yet been officially released, we will leave the study of their association with TDEs for future work. If a certain level of systematic uncertainty is considered, as discussed in Sec.~\ref{sec:sys}, the analysis and conclusions presented in this paper will remain applicable.}}

\section*{Acknowledgements}
This work is supported by the National Key Research and Development Program of China (Grant No. 2022YFF0503304), the National Natural Science Foundation of China (12373042, U1938201,12494573), the Programme of Bagui Scholars Programme (WXG) and Innovation Project of Guangxi Graduate Education (YCBZ2024060).
\label{lastpage}

\bibliography{master}{}

\begin{thebibliography}{}
\expandafter\ifx\csname natexlab\endcsname\relax\def\natexlab#1{#1}\fi
\providecommand{\url}[1]{\href{#1}{#1}}
\providecommand{\dodoi}[1]{doi:~\href{http://doi.org/#1}{\nolinkurl{#1}}}
\providecommand{\doeprint}[1]{\href{http://ascl.net/#1}{\nolinkurl{http://ascl.net/#1}}}
\providecommand{\doarXiv}[1]{\href{https://arxiv.org/abs/#1}{\nolinkurl{https://arxiv.org/abs/#1}}}

\bibitem[{Aartsen {et~al.}(2013{\natexlab{a}})}]{Aartsen:2013jdh}
Aartsen, M.~G., {et~al.} 2013{\natexlab{a}}, Science, 342, 1242856, \dodoi{10.1126/science.1242856}

\bibitem[{Aartsen {et~al.}(2013{\natexlab{b}})}]{Aartsen:2013uuv}
---. 2013{\natexlab{b}}, Astrophys. J., 779, 132, \dodoi{10.1088/0004-637X/779/2/132}

\bibitem[{Aartsen {et~al.}(2013{\natexlab{c}})}]{IceCube:2013cdw}
---. 2013{\natexlab{c}}, Phys. Rev. Lett., 111, 021103, \dodoi{10.1103/PhysRevLett.111.021103}

\bibitem[{Aartsen {et~al.}(2014{\natexlab{a}})}]{IceCube:2014stg}
---. 2014{\natexlab{a}}, Phys. Rev. Lett., 113, 101101, \dodoi{10.1103/PhysRevLett.113.101101}

\bibitem[{Aartsen {et~al.}(2014{\natexlab{b}})}]{Aartsen:2014cva}
---. 2014{\natexlab{b}}, Astrophys. J., 796, 109, \dodoi{10.1088/0004-637X/796/2/109}

\bibitem[{Aartsen {et~al.}(2014{\natexlab{c}})}]{IceCube:2014gqr}
---. 2014{\natexlab{c}}.
\newblock \doarXiv{1412.5106}

\bibitem[{Aartsen {et~al.}(2015)}]{Aartsen:2015knd}
---. 2015, Astrophys. J., 809, 98, \dodoi{10.1088/0004-637X/809/1/98}

\bibitem[{Aartsen {et~al.}(2016)}]{IceCube:2016umi}
---. 2016, Astrophys. J., 833, 3, \dodoi{10.3847/0004-637X/833/1/3}

\bibitem[{Aartsen {et~al.}(2017{\natexlab{a}})}]{Aartsen:2017kpd}
---. 2017{\natexlab{a}}, Nature, 551, 596, \dodoi{10.1038/nature24459}

\bibitem[{Aartsen {et~al.}(2017{\natexlab{b}})}]{IceCube:2016tpw}
---. 2017{\natexlab{b}}, Astrophys. J., 835, 151, \dodoi{10.3847/1538-4357/835/2/151}

\bibitem[{Aartsen {et~al.}(2017{\natexlab{c}})}]{IceCube:2017der}
---. 2017{\natexlab{c}}, Astrophys. J., 846, 136, \dodoi{10.3847/1538-4357/aa8508}

\bibitem[{Aartsen {et~al.}(2019{\natexlab{a}})}]{Aartsen:2018ywr}
---. 2019{\natexlab{a}}, Eur. Phys. J. C, 79, 234, \dodoi{10.1140/epjc/s10052-019-6680-0}

\bibitem[{Aartsen {et~al.}(2019{\natexlab{b}})}]{IceCube:2019lzm}
---. 2019{\natexlab{b}}, Astrophys. J., 886, 12, \dodoi{10.3847/1538-4357/ab4ae2}

\bibitem[{Aartsen {et~al.}(2020{\natexlab{a}})}]{Aartsen:2020aqd}
---. 2020{\natexlab{a}}, Phys. Rev. Lett., 125, 121104, \dodoi{10.1103/PhysRevLett.125.121104}

\bibitem[{Aartsen {et~al.}(2020{\natexlab{b}})}]{Aartsen:2019fau}
---. 2020{\natexlab{b}}, Phys. Rev. Lett., 124, 051103, \dodoi{10.1103/PhysRevLett.124.051103}

\bibitem[{Aartsen {et~al.}(2020{\natexlab{c}})}]{IceCube:2020svz}
---. 2020{\natexlab{c}}, Astrophys. J., 898, 117, \dodoi{10.3847/1538-4357/ab9fa0}

\bibitem[{Aartsen {et~al.}(2021)}]{IceCube:2021rpz}
---. 2021, Nature, 591, 220, \dodoi{10.1038/s41586-021-03256-1}

\bibitem[{Abbasi {et~al.}(2011)}]{Abbasi:2010rd}
Abbasi, R., {et~al.} 2011, Astrophys. J., 732, 18, \dodoi{10.1088/0004-637X/732/1/18}

\bibitem[{Abbasi {et~al.}(2021{\natexlab{a}})}]{Abbasi:2020jmh}
---. 2021{\natexlab{a}}, Phys. Rev. D, 104, 022002, \dodoi{10.1103/PhysRevD.104.022002}

\bibitem[{Abbasi {et~al.}(2021{\natexlab{b}})}]{Abbasi:2021ryj}
---. 2021{\natexlab{b}}, JINST, 16, P07041, \dodoi{10.1088/1748-0221/16/07/P07041}

\bibitem[{Abbasi {et~al.}(2022{\natexlab{a}})}]{IceCube:2021uhz}
---. 2022{\natexlab{a}}, Astrophys. J., 928, 50, \dodoi{10.3847/1538-4357/ac4d29}

\bibitem[{Abbasi {et~al.}(2022{\natexlab{b}})}]{IceCube:2022der}
---. 2022{\natexlab{b}}, Science, 378, 538, \dodoi{10.1126/science.abg3395}

\bibitem[{Abbasi {et~al.}(2023{\natexlab{a}})}]{IceCube:2023agq}
---. 2023{\natexlab{a}}, Astrophys. J. Suppl., 269, 25, \dodoi{10.3847/1538-4365/acfa95}

\bibitem[{Abbasi {et~al.}(2023{\natexlab{b}})}]{IceCube:2023htm}
---. 2023{\natexlab{b}}, Astrophys. J., 954, 75, \dodoi{10.3847/1538-4357/acdfcb}

\bibitem[{Ackermann {et~al.}(2022)}]{Ackermann:2022rqc}
Ackermann, M., {et~al.} 2022, JHEAp, 36, 55, \dodoi{10.1016/j.jheap.2022.08.001}

\bibitem[{Adrian-Martinez {et~al.}(2016{\natexlab{a}})}]{ANTARES:2015moa}
Adrian-Martinez, S., {et~al.} 2016{\natexlab{a}}, Astrophys. J., 823, 65, \dodoi{10.3847/0004-637X/823/1/65}

\bibitem[{Adrian-Martinez {et~al.}(2016{\natexlab{b}})}]{KM3Net:2016zxf}
---. 2016{\natexlab{b}}, J. Phys. G, 43, 084001, \dodoi{10.1088/0954-3899/43/8/084001}

\bibitem[{Aguilar {et~al.}(2021)}]{RNO-G:2020rmc}
Aguilar, J.~A., {et~al.} 2021, JINST, 16, P03025, \dodoi{10.1088/1748-0221/16/03/P03025}

\bibitem[{Ahlers \& Halzen(2014)}]{Ahlers:2014ioa}
Ahlers, M., \& Halzen, F. 2014, Phys. Rev. D, 90, 043005, \dodoi{10.1103/PhysRevD.90.043005}

\bibitem[{Avrorin {et~al.}(2018)}]{Baikal-GVD:2018isr}
Avrorin, A.~D., {et~al.} 2018, EPJ Web Conf., 191, 01006, \dodoi{10.1051/epjconf/201819101006}

\bibitem[{Bartos {et~al.}(2017)Bartos, Ahrens, Finley, \& Marka}]{Bartos:2016wud}
Bartos, I., Ahrens, M., Finley, C., \& Marka, S. 2017, Phys. Rev. D, 96, 023003, \dodoi{10.1103/PhysRevD.96.023003}

\bibitem[{Bartos {et~al.}(2019)Bartos, Veske, Keivani, Marka, Countryman, Blaufuss, Finley, \& Marka}]{Bartos:2018jco}
Bartos, I., Veske, D., Keivani, A., {et~al.} 2019, Phys. Rev. D, 100, 083017, \dodoi{10.1103/PhysRevD.100.083017}

\bibitem[{Beacom \& Kistler(2007)}]{Beacom:2007yu}
Beacom, J.~F., \& Kistler, M.~D. 2007, Phys. Rev. D, 75, 083001, \dodoi{10.1103/PhysRevD.75.083001}

\bibitem[{Bechtol {et~al.}(2017)Bechtol, Ahlers, Di~Mauro, Ajello, \& Vandenbroucke}]{Bechtol:2015uqb}
Bechtol, K., Ahlers, M., Di~Mauro, M., Ajello, M., \& Vandenbroucke, J. 2017, Astrophys. J., 836, 47, \dodoi{10.3847/1538-4357/836/1/47}

\bibitem[{Bellenghi {et~al.}(2023)Bellenghi, Padovani, Resconi, \& Giommi}]{Bellenghi:2023yza}
Bellenghi, C., Padovani, P., Resconi, E., \& Giommi, P. 2023, Astrophys. J. Lett., 955, L32, \dodoi{10.3847/2041-8213/acf711}

\bibitem[{{Bellm} {et~al.}(2019){Bellm}, {Kulkarni}, {Graham}, {Dekany}, {Smith}, {Riddle}, {Masci}, {Helou}, {Prince}, {Adams}, {Barbarino}, {Barlow}, {Bauer}, {Beck}, {Belicki}, {Biswas}, {Blagorodnova}, {Bodewits}, {Bolin}, {Brinnel}, {Brooke}, {Bue}, {Bulla}, {Burruss}, {Cenko}, {Chang}, {Connolly}, {Coughlin}, {Cromer}, {Cunningham}, {De}, {Delacroix}, {Desai}, {Duev}, {Eadie}, {Farnham}, {Feeney}, {Feindt}, {Flynn}, {Franckowiak}, {Frederick}, {Fremling}, {Gal-Yam}, {Gezari}, {Giomi}, {Goldstein}, {Golkhou}, {Goobar}, {Groom}, {Hacopians}, {Hale}, {Henning}, {Ho}, {Hover}, {Howell}, {Hung}, {Huppenkothen}, {Imel}, {Ip}, {Ivezi{\'c}}, {Jackson}, {Jones}, {Juric}, {Kasliwal}, {Kaspi}, {Kaye}, {Kelley}, {Kowalski}, {Kramer}, {Kupfer}, {Landry}, {Laher}, {Lee}, {Lin}, {Lin}, {Lunnan}, {Giomi}, {Mahabal}, {Mao}, {Miller}, {Monkewitz}, {Murphy}, {Ngeow}, {Nordin}, {Nugent}, {Ofek}, {Patterson}, {Penprase}, {Porter}, {Rauch}, {Rebbapragada}, {Reiley}, {Rigault}, {Rodriguez}, {van Roestel}, {Rusholme}, {van
  Santen}, {Schulze}, {Shupe}, {Singer}, {Soumagnac}, {Stein}, {Surace}, {Sollerman}, {Szkody}, {Taddia}, {Terek}, {Van Sistine}, {van Velzen}, {Vestrand}, {Walters}, {Ward}, {Ye}, {Yu}, {Yan}, \& {Zolkower}}]{2019PASP..131a8002B}
{Bellm}, E.~C., {Kulkarni}, S.~R., {Graham}, M.~J., {et~al.} 2019, \pasp, 131, 018002, \dodoi{10.1088/1538-3873/aaecbe}

\bibitem[{{Bricman} \& {Gomboc}(2020)}]{2020ApJ...890...73B}
{Bricman}, K., \& {Gomboc}, A. 2020, \apj, 890, 73, \dodoi{10.3847/1538-4357/ab6989}

\bibitem[{Buson {et~al.}(2022)}]{Buson:2022fyf}
Buson, S., {et~al.} 2022, Astrophys. J. Lett., 933, L43, \dodoi{10.3847/2041-8213/ac7d5b}

\bibitem[{Buson {et~al.}(2023)Buson, Tramacere, Oswald, Barbano, de~Clairfontaine, Pfeiffer, Azzollini, Baghmanyan, \& Ajello}]{Buson:2023irp}
Buson, S., Tramacere, A., Oswald, L., {et~al.} 2023.
\newblock \doarXiv{2305.11263}

\bibitem[{Bustamante \& Ahlers(2019)}]{Bustamante:2019sdb}
Bustamante, M., \& Ahlers, M. 2019, Phys. Rev. Lett., 122, 241101, \dodoi{10.1103/PhysRevLett.122.241101}

\bibitem[{Bustamante \& Connolly(2019)}]{Bustamante:2017xuy}
Bustamante, M., \& Connolly, A. 2019, Phys. Rev. Lett., 122, 041101, \dodoi{10.1103/PhysRevLett.122.041101}

\bibitem[{Cendes {et~al.}(2021)Cendes, Alexander, Berger, Eftekhari, Williams, \& Chornock}]{Cendes:2021bvp}
Cendes, Y., Alexander, K.~D., Berger, E., {et~al.} 2021, Astrophys. J., 919, 127, \dodoi{10.3847/1538-4357/ac110a}

\bibitem[{Chang {et~al.}(2024)Chang, Zhou, Murase, \& Kamionkowski}]{Chang:2022hqj}
Chang, P.-W., Zhou, B., Murase, K., \& Kamionkowski, M. 2024, Phys. Rev. D, 109, 103041, \dodoi{10.1103/PhysRevD.109.103041}

\bibitem[{Dai \& Fang(2017)}]{Dai:2016gtz}
Dai, L., \& Fang, K. 2017, Mon. Not. Roy. Astron. Soc., 469, 1354, \dodoi{10.1093/mnras/stx863}

\bibitem[{Dai {et~al.}(2015)Dai, McKinney, \& Miller}]{Dai:2015eua}
Dai, L., McKinney, J.~C., \& Miller, M.~C. 2015, Astrophys. J. Lett., 812, L39, \dodoi{10.1088/2041-8205/812/2/L39}

\bibitem[{Davison \& Hinkley(1997)}]{Davison_Hinkley_1997}
Davison, A.~C., \& Hinkley, D.~V. 1997, Bootstrap Methods and their Application, Cambridge Series in Statistical and Probabilistic Mathematics (Cambridge University Press)

\bibitem[{Fang {et~al.}(2020)Fang, Metzger, Vurm, Aydi, \& Chomiuk}]{Fang:2020bkm}
Fang, K., Metzger, B.~D., Vurm, I., Aydi, E., \& Chomiuk, L. 2020, Astrophys. J., 904, 4, \dodoi{10.3847/1538-4357/abbc6e}

\bibitem[{Fang {et~al.}(2018)}]{Fang:2017mhl}
Fang, K., {et~al.} 2018, PoS, ICRC2017, 996, \dodoi{10.22323/1.301.0996}

\bibitem[{Frederick {et~al.}(2021)}]{Frederick:2020sjz}
Frederick, S., {et~al.} 2021, Astrophys. J., 920, 56, \dodoi{10.3847/1538-4357/ac110f}

\bibitem[{Fremling {et~al.}(2020)}]{Fremling:2019dvl}
Fremling, U.~C., {et~al.} 2020, Astrophys. J., 895, 32, \dodoi{10.3847/1538-4357/ab8943}

\bibitem[{Gonzalez-Garcia {et~al.}(2005)Gonzalez-Garcia, Halzen, \& Maltoni}]{GonzalezGarcia:2005xw}
Gonzalez-Garcia, M.~C., Halzen, F., \& Maltoni, M. 2005, Phys. Rev. D, 71, 093010, \dodoi{10.1103/PhysRevD.71.093010}

\bibitem[{Goodwin {et~al.}(2022)Goodwin, van Velzen, Miller-Jones, Mummery, Bietenholz, Wederfoort, Hammerstein, Bonnerot, Hoffmann, \& Yan}]{Goodwin:2022zik}
Goodwin, A.~J., van Velzen, S., Miller-Jones, J. C.~A., {et~al.} 2022, Mon. Not. Roy. Astron. Soc., 511, 5328, \dodoi{10.1093/mnras/stac333}

\bibitem[{Graham {et~al.}(2019)}]{Graham:2019qsw}
Graham, M.~J., {et~al.} 2019, Publ. Astron. Soc. Pac., 131, 078001, \dodoi{10.1088/1538-3873/ab006c}

\bibitem[{Guo {et~al.}(2023)Guo, L\"u, Huang, Li, Zhu, \& Liang}]{Guo:2023axz}
Guo, X.-K., L\"u, Y.-F., Huang, Y.-B., {et~al.} 2023, Phys. Rev. D, 108, 043001, \dodoi{10.1103/PhysRevD.108.043001}

\bibitem[{{Guolo} {et~al.}(2024){Guolo}, {Gezari}, {Yao}, {van Velzen}, {Hammerstein}, {Cenko}, \& {Tokayer}}]{2024ApJ...966..160G}
{Guolo}, M., {Gezari}, S., {Yao}, Y., {et~al.} 2024, \apj, 966, 160, \dodoi{10.3847/1538-4357/ad2f9f}

\bibitem[{Hayasaki(2021)}]{Hayasaki:2021jem}
Hayasaki, K. 2021, \dodoi{10.1038/s41550-021-01309-z}

\bibitem[{Hayasaki \& Yamazaki(2019)}]{Hayasaki:2019kjy}
Hayasaki, K., \& Yamazaki, R. 2019, \dodoi{10.3847/1538-4357/ab44ca}

\bibitem[{Hinkle {et~al.}(2020)}]{Hinkle:2020ilv}
Hinkle, J.~T., {et~al.} 2020, Mon. Not. Roy. Astron. Soc., 500, 1673, \dodoi{10.1093/mnras/staa3170}

\bibitem[{Hovatta {et~al.}(2021)}]{Hovatta:2020lor}
Hovatta, T., {et~al.} 2021, Astron. Astrophys., 650, A83, \dodoi{10.1051/0004-6361/202039481}

\bibitem[{Ioka \& Murase(2014)}]{Ioka:2014kca}
Ioka, K., \& Murase, K. 2014, PTEP, 2014, 061E01, \dodoi{10.1093/ptep/ptu090}

\bibitem[{{Ivezi{\'c}} {et~al.}(2019){Ivezi{\'c}}, {Kahn}, {Tyson}, {Abel}, {Acosta}, {Allsman}, {Alonso}, {AlSayyad}, {Anderson}, {Andrew}, {Angel}, {Angeli}, {Ansari}, {Antilogus}, {Araujo}, {Armstrong}, {Arndt}, {Astier}, {Aubourg}, {Auza}, {Axelrod}, {Bard}, {Barr}, {Barrau}, {Bartlett}, {Bauer}, {Bauman}, {Baumont}, {Bechtol}, {Bechtol}, {Becker}, {Becla}, {Beldica}, {Bellavia}, {Bianco}, {Biswas}, {Blanc}, {Blazek}, {Blandford}, {Bloom}, {Bogart}, {Bond}, {Booth}, {Borgland}, {Borne}, {Bosch}, {Boutigny}, {Brackett}, {Bradshaw}, {Brandt}, {Brown}, {Bullock}, {Burchat}, {Burke}, {Cagnoli}, {Calabrese}, {Callahan}, {Callen}, {Carlin}, {Carlson}, {Chandrasekharan}, {Charles-Emerson}, {Chesley}, {Cheu}, {Chiang}, {Chiang}, {Chirino}, {Chow}, {Ciardi}, {Claver}, {Cohen-Tanugi}, {Cockrum}, {Coles}, {Connolly}, {Cook}, {Cooray}, {Covey}, {Cribbs}, {Cui}, {Cutri}, {Daly}, {Daniel}, {Daruich}, {Daubard}, {Daues}, {Dawson}, {Delgado}, {Dellapenna}, {de Peyster}, {de Val-Borro}, {Digel}, {Doherty}, {Dubois},
  {Dubois-Felsmann}, {Durech}, {Economou}, {Eifler}, {Eracleous}, {Emmons}, {Fausti Neto}, {Ferguson}, {Figueroa}, {Fisher-Levine}, {Focke}, {Foss}, {Frank}, {Freemon}, {Gangler}, {Gawiser}, {Geary}, {Gee}, {Geha}, {Gessner}, {Gibson}, {Gilmore}, {Glanzman}, {Glick}, {Goldina}, {Goldstein}, {Goodenow}, {Graham}, {Gressler}, {Gris}, {Guy}, {Guyonnet}, {Haller}, {Harris}, {Hascall}, {Haupt}, {Hernandez}, {Herrmann}, {Hileman}, {Hoblitt}, {Hodgson}, {Hogan}, {Howard}, {Huang}, {Huffer}, {Ingraham}, {Innes}, {Jacoby}, {Jain}, {Jammes}, {Jee}, {Jenness}, {Jernigan}, {Jevremovi{\'c}}, {Johns}, {Johnson}, {Johnson}, {Jones}, {Juramy-Gilles}, {Juri{\'c}}, {Kalirai}, {Kallivayalil}, {Kalmbach}, {Kantor}, {Karst}, {Kasliwal}, {Kelly}, {Kessler}, {Kinnison}, {Kirkby}, {Knox}, {Kotov}, {Krabbendam}, {Krughoff}, {Kub{\'a}nek}, {Kuczewski}, {Kulkarni}, {Ku}, {Kurita}, {Lage}, {Lambert}, {Lange}, {Langton}, {Le Guillou}, {Levine}, {Liang}, {Lim}, {Lintott}, {Long}, {Lopez}, {Lotz}, {Lupton}, {Lust}, {MacArthur}, {Mahabal},
  {Mandelbaum}, {Markiewicz}, {Marsh}, {Marshall}, {Marshall}, {May}, {McKercher}, {McQueen}, {Meyers}, {Migliore}, {Miller}, \& {Mills}}]{2019ApJ...873..111I}
{Ivezi{\'c}}, {\v{Z}}., {Kahn}, S.~M., {Tyson}, J.~A., {et~al.} 2019, \apj, 873, 111, \dodoi{10.3847/1538-4357/ab042c}

\bibitem[{Jiang {et~al.}(2016)Jiang, Dou, Wang, Yang, Lyu, \& Zhou}]{Jiang:2016znw}
Jiang, N., Dou, L., Wang, T., {et~al.} 2016, Astrophys. J. Lett., 828, L14, \dodoi{10.3847/2041-8205/828/1/L14}

\bibitem[{Jiang {et~al.}(2023)Jiang, Zhou, Zhu, Wang, \& Wang}]{Jiang:2023kbb}
Jiang, N., Zhou, Z., Zhu, J., Wang, Y., \& Wang, T. 2023, Astrophys. J. Lett., 953, L12, \dodoi{10.3847/2041-8213/acebe3}

\bibitem[{{Jiang} {et~al.}(2021){Jiang}, {Wang}, {Dou}, {Shu}, {Hu}, {Liu}, {Wang}, {Yan}, {Sheng}, {Yang}, {Sun}, \& {Zhou}}]{2021ApJS..252...32J}
{Jiang}, N., {Wang}, T., {Dou}, L., {et~al.} 2021, \apjs, 252, 32, \dodoi{10.3847/1538-4365/abd1dc}

\bibitem[{Kistler \& Beacom(2006)}]{Kistler:2006hp}
Kistler, M.~D., \& Beacom, J.~F. 2006, Phys. Rev. D, 74, 063007, \dodoi{10.1103/PhysRevD.74.063007}

\bibitem[{Kistler \& Laha(2018)}]{Kistler:2016ask}
Kistler, M.~D., \& Laha, R. 2018, Phys. Rev. Lett., 120, 241105, \dodoi{10.1103/PhysRevLett.120.241105}

\bibitem[{Kouch {et~al.}(2024)}]{Kouch:2024xtd}
Kouch, P.~M., {et~al.} 2024, Astron. Astrophys., 690, A111, \dodoi{10.1051/0004-6361/202347624}

\bibitem[{Kovalev {et~al.}(2022)Kovalev, Plavin, \& Troitsky}]{Kovalev:2022izi}
Kovalev, Y.~Y., Plavin, A.~V., \& Troitsky, S.~V. 2022, Astrophys. J. Lett., 940, L41, \dodoi{10.3847/2041-8213/aca1ae}

\bibitem[{Li {et~al.}(2024)Li, Yuan, He, Wang, Zhu, Liang, Jiang, \& Wei}]{Li:2024qcp}
Li, R.-L., Yuan, C., He, H.-N., {et~al.} 2024.
\newblock \doarXiv{2411.06440}

\bibitem[{{Lin} {et~al.}(2022){Lin}, {Jiang}, \& {Kong}}]{2022MNRAS.513.2422L}
{Lin}, Z., {Jiang}, N., \& {Kong}, X. 2022, \mnras, 513, 2422, \dodoi{10.1093/mnras/stac946}

\bibitem[{Lin {et~al.}(2022)Lin, Jiang, Kong, Huang, Lin, Zhu, \& Wang}]{Lin:2022jvw}
Lin, Z., Jiang, N., Kong, X., {et~al.} 2022, Astrophys. J. Lett., 939, L33, \dodoi{10.3847/2041-8213/ac9c63}

\bibitem[{Liu {et~al.}(2020)Liu, Xi, \& Wang}]{Liu:2020isi}
Liu, R.-Y., Xi, S.-Q., \& Wang, X.-Y. 2020, Phys. Rev. D, 102, 083028, \dodoi{10.1103/PhysRevD.102.083028}

\bibitem[{Liu {et~al.}(2022)Liu, Dou, Chen, \& Shen}]{Liu:2019dnt}
Liu, X.-L., Dou, L.-M., Chen, J.-H., \& Shen, R.-F. 2022, Astrophys. J., 925, 67, \dodoi{10.3847/1538-4357/ac33a9}

\bibitem[{Lu {et~al.}(2024{\natexlab{a}})Lu, Liang, Ouyang, Li, \& Wang}]{Lu:2024flp}
Lu, M.-X., Liang, Y.-F., Ouyang, X., Li, R.-L., \& Wang, X.-G. 2024{\natexlab{a}}.
\newblock \doarXiv{2404.19730}

\bibitem[{Lu {et~al.}(2016)Lu, Kumar, \& Evans}]{Lu:2015jfg}
Lu, W., Kumar, P., \& Evans, N.~J. 2016, Mon. Not. Roy. Astron. Soc., 458, 575, \dodoi{10.1093/mnras/stw307}

\bibitem[{Lu {et~al.}(2024{\natexlab{b}})Lu, Matsumoto, \& Matzner}]{Lu:2023miv}
Lu, W., Matsumoto, T., \& Matzner, C.~D. 2024{\natexlab{b}}, Mon. Not. Roy. Astron. Soc., 533, 979, \dodoi{10.1093/mnras/stae1770}

\bibitem[{L\"u {et~al.}(2024)L\"u, Zhu, Li, Guo, Liu, Huang, \& Liang}]{Lu:2024jbq}
L\"u, Y.-F., Zhu, B.-Y., Li, R.-L., {et~al.} 2024, Res. Astron. Astrophys., 24, 035008, \dodoi{10.1088/1674-4527/ad204e}

\bibitem[{Lunardini \& Winter(2017)}]{Lunardini:2016xwi}
Lunardini, C., \& Winter, W. 2017, Phys. Rev. D, 95, 123001, \dodoi{10.1103/PhysRevD.95.123001}

\bibitem[{{Mainzer} {et~al.}(2014){Mainzer}, {Bauer}, {Cutri}, {Grav}, {Masiero}, {Beck}, {Clarkson}, {Conrow}, {Dailey}, {Eisenhardt}, {Fabinsky}, {Fajardo-Acosta}, {Fowler}, {Gelino}, {Grillmair}, {Heinrichsen}, {Kendall}, {Kirkpatrick}, {Liu}, {Masci}, {McCallon}, {Nugent}, {Papin}, {Rice}, {Royer}, {Ryan}, {Sevilla}, {Sonnett}, {Stevenson}, {Thompson}, {Wheelock}, {Wiemer}, {Wittman}, {Wright}, \& {Yan}}]{2014ApJ...792...30M}
{Mainzer}, A., {Bauer}, J., {Cutri}, R.~M., {et~al.} 2014, \apj, 792, 30, \dodoi{10.1088/0004-637X/792/1/30}

\bibitem[{{Mainzer} {et~al.}(2023){Mainzer}, {Masiero}, {Abell}, {Bauer}, {Bottke}, {Buratti}, {Carey}, {Cotto-Figueroa}, {Cutri}, {Dahlen}, {Eisenhardt}, {Fernandez}, {Furfaro}, {Grav}, {Hoffman}, {Kelley}, {Kim}, {Kirkpatrick}, {Lawler}, {Lilly}, {Liu}, {Marocco}, {Marsh}, {Masci}, {McMurtry}, {Pourrahmani}, {Reinhart}, {Ressler}, {Satpathy}, {Schambeau}, {Sonnett}, {Spahr}, {Surace}, {Vaquero}, {Wright}, {Zengilowski}, \& {NEO Surveyor Mission Team}}]{2023PSJ.....4..224M}
{Mainzer}, A.~K., {Masiero}, J.~R., {Abell}, P.~A., {et~al.} 2023, Planet. Sci. J., 4, 224, \dodoi{10.3847/PSJ/ad0468}

\bibitem[{Masci {et~al.}(2018)}]{Masci:2018neq}
Masci, F.~J., {et~al.} 2018, Publ. Astron. Soc. Pac., 131, 018003, \dodoi{10.1088/1538-3873/aae8ac}

\bibitem[{Mou {et~al.}(2021)}]{Mou:2020csb}
Mou, G., {et~al.} 2021, Astrophys. J., 908, 197, \dodoi{10.3847/1538-4357/abd475}

\bibitem[{Mukhopadhyay {et~al.}(2024)Mukhopadhyay, Bhattacharya, \& Murase}]{Mukhopadhyay:2023mld}
Mukhopadhyay, M., Bhattacharya, M., \& Murase, K. 2024, Mon. Not. Roy. Astron. Soc., 534, 1528, \dodoi{10.1093/mnras/stae2080}

\bibitem[{Murase {et~al.}(2013)Murase, Ahlers, \& Lacki}]{Murase:2013rfa}
Murase, K., Ahlers, M., \& Lacki, B.~C. 2013, Phys. Rev. D, 88, 121301, \dodoi{10.1103/PhysRevD.88.121301}

\bibitem[{Murase {et~al.}(2014)Murase, Inoue, \& Dermer}]{Murase:2014foa}
Murase, K., Inoue, Y., \& Dermer, C.~D. 2014, Phys. Rev. D, 90, 023007, \dodoi{10.1103/PhysRevD.90.023007}

\bibitem[{Murase \& Ioka(2013)}]{Murase:2013ffa}
Murase, K., \& Ioka, K. 2013, Phys. Rev. Lett., 111, 121102, \dodoi{10.1103/PhysRevLett.111.121102}

\bibitem[{Murase {et~al.}(2020)Murase, Kimura, Zhang, Oikonomou, \& Petropoulou}]{Murase:2020lnu}
Murase, K., Kimura, S.~S., Zhang, B.~T., Oikonomou, F., \& Petropoulou, M. 2020, Astrophys. J., 902, 108, \dodoi{10.3847/1538-4357/abb3c0}

\bibitem[{Murase {et~al.}(2011)Murase, Thompson, Lacki, \& Beacom}]{Murase:2010cu}
Murase, K., Thompson, T.~A., Lacki, B.~C., \& Beacom, J.~F. 2011, Phys. Rev. D, 84, 043003, \dodoi{10.1103/PhysRevD.84.043003}

\bibitem[{{NEOWISE Team}(2020)}]{https://doi.org/10.26131/irsa144}
{NEOWISE Team}. 2020, NEOWISE-R Single Exposure (L1b) Source Table,  IPAC, \dodoi{10.26131/IRSA144}

\bibitem[{Ng \& Beacom(2014)}]{Ng:2014pca}
Ng, K. C.~Y., \& Beacom, J.~F. 2014, Phys. Rev. D, 90, 065035, \dodoi{10.1103/PhysRevD.90.065035}

\bibitem[{Perley {et~al.}(2020)}]{Perley:2020ajb}
Perley, D.~A., {et~al.} 2020, Astrophys. J., 904, 35, \dodoi{10.3847/1538-4357/abbd98}

\bibitem[{Plavin {et~al.}(2020)Plavin, Kovalev, Kovalev, \& Troitsky}]{Plavin:2020emb}
Plavin, A., Kovalev, Y.~Y., Kovalev, Y.~A., \& Troitsky, S. 2020, Astrophys. J., 894, 101, \dodoi{10.3847/1538-4357/ab86bd}

\bibitem[{Plavin {et~al.}(2024)Plavin, Burenin, Kovalev, Lutovinov, Starobinsky, Troitsky, \& Zakharov}]{Plavin:2023wsb}
Plavin, A.~V., Burenin, R.~A., Kovalev, Y.~Y., {et~al.} 2024, JCAP, 05, 133, \dodoi{10.1088/1475-7516/2024/05/133}

\bibitem[{Plavin {et~al.}(2023)Plavin, Kovalev, Kovalev, \& Troitsky}]{Plavin:2022oyy}
Plavin, A.~V., Kovalev, Y.~Y., Kovalev, Y.~A., \& Troitsky, S.~V. 2023, Mon. Not. Roy. Astron. Soc., 523, 1799, \dodoi{10.1093/mnras/stad1467}

\bibitem[{Plestid \& Zhou(2024)}]{Plestid:2024bva}
Plestid, R., \& Zhou, B. 2024.
\newblock \doarXiv{2403.07984}

\bibitem[{{Rees}(1988)}]{1988Natur.333..523R}
{Rees}, M.~J. 1988, \nat, 333, 523, \dodoi{10.1038/333523a0}

\bibitem[{Reusch {et~al.}(2022)}]{Reusch:2021ztx}
Reusch, S., {et~al.} 2022, Phys. Rev. Lett., 128, 221101, \dodoi{10.1103/PhysRevLett.128.221101}

\bibitem[{Senno {et~al.}(2017)Senno, Murase, \& Meszaros}]{Senno:2016bso}
Senno, N., Murase, K., \& Meszaros, P. 2017, Astrophys. J., 838, 3, \dodoi{10.3847/1538-4357/aa6344}

\bibitem[{{Stein} {et~al.}(2021){Stein}, {van Velzen}, {Kowalski}, {Franckowiak}, {Gezari}, {Miller-Jones}, {Frederick}, {Sfaradi}, {Bietenholz}, {Horesh}, {Fender}, {Garrappa}, {Ahumada}, {Andreoni}, {Belicki}, {Bellm}, {B{\"o}ttcher}, {Brinnel}, {Burruss}, {Cenko}, {Coughlin}, {Cunningham}, {Drake}, {Farrar}, {Feeney}, {Foley}, {Gal-Yam}, {Golkhou}, {Goobar}, {Graham}, {Hammerstein}, {Helou}, {Hung}, {Kasliwal}, {Kilpatrick}, {Kong}, {Kupfer}, {Laher}, {Mahabal}, {Masci}, {Necker}, {Nordin}, {Perley}, {Rigault}, {Reusch}, {Rodriguez}, {Rojas-Bravo}, {Rusholme}, {Shupe}, {Singer}, {Sollerman}, {Soumagnac}, {Stern}, {Taggart}, {van Santen}, {Ward}, {Woudt}, \& {Yao}}]{2021NatAs...5..510S}
{Stein}, R., {van Velzen}, S., {Kowalski}, M., {et~al.} 2021, Nature Astronomy, 5, 510, \dodoi{10.1038/s41550-020-01295-8}

\bibitem[{Stein {et~al.}(2021)}]{Stein:2020xhk}
Stein, R., {et~al.} 2021, Nature Astron., 5, 510, \dodoi{10.1038/s41550-020-01295-8}

\bibitem[{Sudoh {et~al.}(2018)Sudoh, Totani, \& Kawanaka}]{Sudoh:2018ana}
Sudoh, T., Totani, T., \& Kawanaka, N. 2018, Publ. Astron. Soc. Jap., 70, Publications of the Astronomical Society of Japan, Volume 70, Issue 3, 1 June 2018, 49, https://doi.org/10.1093/pasj/psy039, \dodoi{10.1093/pasj/psy039}

\bibitem[{Suray \& Troitsky(2023)}]{Suray:2023lsa}
Suray, A., \& Troitsky, S. 2023, Mon. Not. Roy. Astron. Soc., 527, L26, \dodoi{10.1093/mnrasl/slad136}

\bibitem[{Tamborra {et~al.}(2014)Tamborra, Ando, \& Murase}]{Tamborra:2014xia}
Tamborra, I., Ando, S., \& Murase, K. 2014, JCAP, 09, 043, \dodoi{10.1088/1475-7516/2014/09/043}

\bibitem[{Trakhtenbrot {et~al.}(2019)}]{Trakhtenbrot:2019seb}
Trakhtenbrot, B., {et~al.} 2019, Nature Astron., 3, 242, \dodoi{10.1038/s41550-018-0661-3}

\bibitem[{van Velzen {et~al.}(2020)van Velzen, Holoien, Onori, Hung, \& Arcavi}]{vanVelzen:2020hrd}
van Velzen, S., Holoien, T. W.~S., Onori, F., Hung, T., \& Arcavi, I. 2020, Space Sci. Rev., 216, 124, \dodoi{10.1007/s11214-020-00753-z}

\bibitem[{van Velzen {et~al.}(2016)van Velzen, Mendez, Krolik, \& Gorjian}]{vanVelzen:2016jsk}
van Velzen, S., Mendez, A.~J., Krolik, J.~H., \& Gorjian, V. 2016, Astrophys. J., 829, 19, \dodoi{10.3847/0004-637X/829/1/19}

\bibitem[{van Velzen {et~al.}(2024)}]{vanVelzen:2021zsm}
van Velzen, S., {et~al.} 2024, Mon. Not. Roy. Astron. Soc., 529, 2559, \dodoi{10.1093/mnras/stae610}

\bibitem[{Veres {et~al.}(2024)}]{Veres:2024qcm}
Veres, P.~M., {et~al.} 2024.
\newblock \doarXiv{2408.17419}

\bibitem[{{Wang} {et~al.}(2023){Wang}, {Liu}, {Cai}, {Geng}, {Fang}, {He}, {Jiang}, {Jiang}, {Kong}, {Li}, {Li}, {Luo}, {Pan}, {Wu}, {Yang}, {Yu}, {Zheng}, {Zhu}, {Cai}, {Chen}, {Chen}, {Dai}, {Fan}, {Fan}, {Fang}, {He}, {Hu}, {Hu}, {Jin}, {Jiang}, {Li}, {Li}, {Li}, {Liang}, {Lin}, {Liu}, {Liu}, {Liu}, {Liu}, {Liu}, {Lou}, {Qu}, {Sheng}, {Shi}, {Shu}, {Su}, {Sun}, {Wang}, {Wang}, {Wang}, {Wang}, {Wei}, {Wei}, {Xue}, {Yan}, {Yang}, {Yuan}, {Yuan}, {Zhang}, {Zhang}, {Zhao}, \& {Zhao}}]{2023SCPMA..6609512W}
{Wang}, T., {Liu}, G., {Cai}, Z., {et~al.} 2023, Science China Physics, Mechanics, and Astronomy, 66, 109512, \dodoi{10.1007/s11433-023-2197-5}

\bibitem[{Wang \& Liu(2016)}]{Wang:2015mmh}
Wang, X.-Y., \& Liu, R.-Y. 2016, Phys. Rev. D, 93, 083005, \dodoi{10.1103/PhysRevD.93.083005}

\bibitem[{Wang {et~al.}(2011)Wang, Liu, Dai, \& Cheng}]{Wang:2011ip}
Wang, X.-Y., Liu, R.-Y., Dai, Z.-G., \& Cheng, K.~S. 2011, Phys. Rev. D, 84, 081301, \dodoi{10.1103/PhysRevD.84.081301}

\bibitem[{Winter \& Lunardini(2021)}]{Winter:2020ptf}
Winter, W., \& Lunardini, C. 2021, Nature Astron., 5, 472, \dodoi{10.1038/s41550-021-01343-x}

\bibitem[{Winter \& Lunardini(2023)}]{Winter:2022fpf}
---. 2023, Astrophys. J., 948, 42, \dodoi{10.3847/1538-4357/acbe9e}

\bibitem[{Wiseman {et~al.}(2023)}]{Wiseman:2023tnv}
Wiseman, P., {et~al.} 2023, Mon. Not. Roy. Astron. Soc., 522, 3992, \dodoi{10.1093/mnras/stad1000}

\bibitem[{{Wright} {et~al.}(2010){Wright}, {Eisenhardt}, {Mainzer}, {Ressler}, {Cutri}, {Jarrett}, {Kirkpatrick}, {Padgett}, {McMillan}, {Skrutskie}, {Stanford}, {Cohen}, {Walker}, {Mather}, {Leisawitz}, {Gautier}, {McLean}, {Benford}, {Lonsdale}, {Blain}, {Mendez}, {Irace}, {Duval}, {Liu}, {Royer}, {Heinrichsen}, {Howard}, {Shannon}, {Kendall}, {Walsh}, {Larsen}, {Cardon}, {Schick}, {Schwalm}, {Abid}, {Fabinsky}, {Naes}, \& {Tsai}}]{2010AJ....140.1868W}
{Wright}, E.~L., {Eisenhardt}, P. R.~M., {Mainzer}, A.~K., {et~al.} 2010, \aj, 140, 1868, \dodoi{10.1088/0004-6256/140/6/1868}

\bibitem[{Wu {et~al.}(2022)Wu, Mou, Wang, Wang, \& Li}]{Wu:2021vaw}
Wu, H.-J., Mou, G., Wang, K., Wang, W., \& Li, Z. 2022, Mon. Not. Roy. Astron. Soc., 514, 4406, \dodoi{10.1093/mnras/stac1621}

\bibitem[{Yuan \& Winter(2023)}]{Yuan:2023cmd}
Yuan, C., \& Winter, W. 2023, Astrophys. J., 956, 30, \dodoi{10.3847/1538-4357/acf615}

\bibitem[{Yuan {et~al.}(2024)Yuan, Winter, \& Lunardini}]{Yuan:2024foi}
Yuan, C., Winter, W., \& Lunardini, C. 2024, Astrophys. J., 969, 136, \dodoi{10.3847/1538-4357/ad50a9}

\bibitem[{Zegarelli {et~al.}(2025)Zegarelli, Franckowiak, Sommani, Valtonen-Mattila, \& Yuan}]{zegarelli2025icecat2updatedicecubeevent}
Zegarelli, A., Franckowiak, A., Sommani, G., Valtonen-Mattila, N., \& Yuan, T. 2025, IceCat-2: Updated IceCube Event Catalog of Alert Tracks.
\newblock \doarXiv{2507.06176}

\bibitem[{Zhou \& Beacom(2020{\natexlab{a}})}]{Zhou:2019vxt}
Zhou, B., \& Beacom, J.~F. 2020{\natexlab{a}}, Phys. Rev. D, 101, 036011, \dodoi{10.1103/PhysRevD.101.036011}

\bibitem[{Zhou \& Beacom(2020{\natexlab{b}})}]{Zhou:2019frk}
---. 2020{\natexlab{b}}, Phys. Rev. D, 101, 036010, \dodoi{10.1103/PhysRevD.101.036010}

\bibitem[{Zhou \& Beacom(2022)}]{Zhou:2021xuh}
---. 2022, Phys. Rev. D, 105, 093005, \dodoi{10.1103/PhysRevD.105.093005}

\bibitem[{Zhou {et~al.}(2021)Zhou, Kamionkowski, \& Liang}]{Zhou:2021rhl}
Zhou, B., Kamionkowski, M., \& Liang, Y.-f. 2021, Phys. Rev. D, 103, 123018, \dodoi{10.1103/PhysRevD.103.123018}

\end{thebibliography}
\bibliographystyle{aasjournal}

\end{document}